\documentclass{article}

\usepackage{commath}
\usepackage{graphicx}
\usepackage{fullpage}
\usepackage{subcaption}
\usepackage{units}
\usepackage{hyperref}
\usepackage{siunitx}
\usepackage{listings}
\usepackage{multicol}
\usepackage{bm}
\usepackage{appendix}
\usepackage{textcomp}
\usepackage{gensymb}
\usepackage{booktabs}

\allowdisplaybreaks

\DeclareMathOperator*{\argmin}{argmin}

\title{
	Time stretching of the GeV emission of GRBs: Fermi-LAT data vs geometrical model
}

\author{
	Maxim S. Piskunov$^{1,2}$\thanks{\href{mailto:maxit@ms2.inr.ac.ru}{maxit@ms2.inr.ac.ru}} ~and
	Grigory I. Rubtsov$^{2}$\thanks{\href{mailto:grisha@ms2.inr.ac.ru}{grisha@ms2.inr.ac.ru}} \\
	{\small $^{1}$Faculty of Physics, Moscow State University, Moscow, Russia}\\
	{\small $^{2}$Institute for Nuclear Research of the RAS, Moscow, Russia}
}

\begin{document}
\begin{flushright}  
INR-TH/2014-034
\end{flushright}
\vskip -0.9cm
{\let\newpage\relax \maketitle}

\begin{abstract}
	It is known that the high energy $(> \unit[100]{MeV})$
        emission of gamma-ray bursts is delayed with respect to the
        low energy emission.  However, the dependence of light curves
        on energy has not been studied for the high energy bands.  In
        this paper we consider the bursts observed by Fermi LAT from
        2008 August 4 to 2011 August 1, for which at least $10$
        photons were observed with the energy greater than
        $\unit[1]{GeV}$. These include $4$ bursts: GRB 080916C, GRB
        090510, GRB 090902B, and GRB 090926A.  We use the
        Kolmogorov-Smirnov test to compare the light curves in the two
        bands, $\unit[100]{MeV} < E < \unit[1]{GeV}$ and
        $E > \unit[1]{GeV}$. For GRB 080916C and GRB
        090510 the light curves in the two bands are statistically
        compatible. However, for GRB 090926A, the higher-energy light
        curve is stretched compared to the lower-energy one with a
        statistical significance of $3.3 \sigma$ and, for GRB 090902B,
        on the contrary, the lower-energy curve is stretched with $2.3
        \sigma$ significance. We argue that the observed diversity of
        stretching factors may be explained in a simple geometrical
        model. The model assumes that the jet opening angle depends on
        the emission energy in a way that the most energetic photons
        are radiated near the axis of the jet. All the bursts are
        considered equivalent in their rest frames and the observed
        light curves differ only due to different redshifts and view
        directions. The model conforms to the total burst energy
        constraint and matches the Fermi-LAT observations of the
        fraction of GRBs visible in $\unit[100]{MeV} < E <
        \unit[1]{GeV}$ band, which may be observed at higher
        energies. The model predicts the distribution of observable
        stretching factors, which may be tested in the future
        data. Finally, we propose a method to estimate observer's
        off-axis angle based on the stretching factor and the fraction of the
        high-energy photons. The code for modeling is open source and
        is publicly available on GitHub
        (\url{https://github.com/maxitg/GammaRays}).
\end{abstract}

\section{Introduction and summary}

Gamma-ray bursts (GRBs) are among the most energetic events in the
Universe. Therefore, the studies of the emission mechanism and
phenomenology of GRBs may provide the new knowledge in particle
physics. Moreover, the GRBs are observed from cosmological distances
and therefore bear an imprint of the late-time evolution of the
Universe. An extensive studies of these explosions led to a number of
interesting results, see~\cite{Vianello:2013ela,Gehrels:2013xd} for a
review. In particular, the total energy emitted in gamma-rays during a
burst was found to be similar for the different GRBs within an order
of magnitude~\cite{Bloom:2003wy}. This indicates that the most of the
bursts have similar energetics in their rest frames. Moreover, the
high energy radiation of GRBs is shown to follow the predictions of
the synchrotron radiation scenario~\cite{Wang:2013ptaw}.  Several
observations are related to the temporal variations of spectra. This
way, the spectral lags were found between different low energy bands
\cite{Yi:2005ht} and the very-high-energy radiation was discovered to
be extended relative to the x-ray emission
\cite{Castignani:2014gaa,Lange:2013uh,Vianello:2013ela}. In this paper
we go further and compare the temporal extension of GRB gamma-ray
radiation in the two energy bands $\unit[100]{MeV} < E <
\unit[1]{GeV}$ and $E > \unit[1]{GeV}$ (hereafter, low and high-energy
bands) using the data from the Large Area Telescope (LAT) of the Fermi
satellite~\cite{2009ApJ...697.1071A,Ackermann:2012kna}. 

We start with the location, time and duration of bursts from the
Fermi-LAT Gamma-Ray Burst Catalog \cite{Ackermann:2013zfa}. The
photons during the burst and the photons for 24 hours before the burst
are downloaded from the LAT Data
Server~\footnote{\url{http://fermi.gsfc.nasa.gov/cgi-bin/ssc/LAT/LATDataQuery.cgi}}.
The latter are used for the background estimation in both energy bands
with the technique introduced in \cite{Rubtsov:2011qq}. In order to
have enough statistics we require that at least 10 photons are
detected with the energy greater than $\unit[1]{GeV}$. This leaves us
with the four bursts for the time span of the Fermi-LAT Gamma-Ray
Burst Catalog. Namely, GRB~080916C \cite{Tajima:2009az}, GRB~090510
\cite{Ackermann:2010us}, GRB~090902B \cite{Abdo:2009pg} and
GRB~090926A \cite{Bregeon:2011bu}. For these bursts we stretch the
high energy light curve by the arbitrary factor and compare with the
low energy light curve using the Kolmogorov-Smirnov (KS) test. Within
the $95\%$ confidence level the stretching factor is compatible to $1$
(no stretching) for the GRB 080916C and GRB 090510. For the other two
bursts the stretching factor of 1 is, however, excluded.  For GRB
090902B it should be smaller than 1, so the low energy light curve is
stretched with respect to the high energy one, see
fig. \ref{fig:grb090902B}.  Even more significant deviation from the
stretching factor of 1 in the opposite direction is observed for the
GRB 090926A. The high energy light curve of GRB 090926A is stretched
with respect to the low energy one by a factor of at least 1.99 (see
fig. \ref{fig:grb090926A}).  The stretching of the GRB 090926A high
energy light is found with $3.3\sigma$ significance.  All the observed
stretching factors are summarized in Table
\ref{tab:observationResults}, and the detailed procedure for their
calculation is described in Section \ref{sec:observations}.

We propose that the observed energy dependence of the time profiles
may be explained without introduction of the new effects. We argue
that the stretching may originate from the effects of the jet geometry
called curvature effects. These effects were explored by multiple
authors \cite{Nakamura:2001kd,Shen:2005ea,Shenoy:2013cba}.  These
studies concerned x-ray radiation, and considered the homogeneous
distribution of radiation sources throughout the jet. Our model is
based on the reverse assumption that the highest energy radiators are
concentrated near the axis of the jet and consequently the jet opening
angle depends on energy. The latter is motivated by observations if
all bursts are considered equivalent. In this case the energy
dependence of the jet opening angle resolves the contradiction between
the small fraction of GRBs visible above GeV and the hard
spectrum. 

	\begin{figure}
        \centering
        \includegraphics[width=0.7\textwidth]{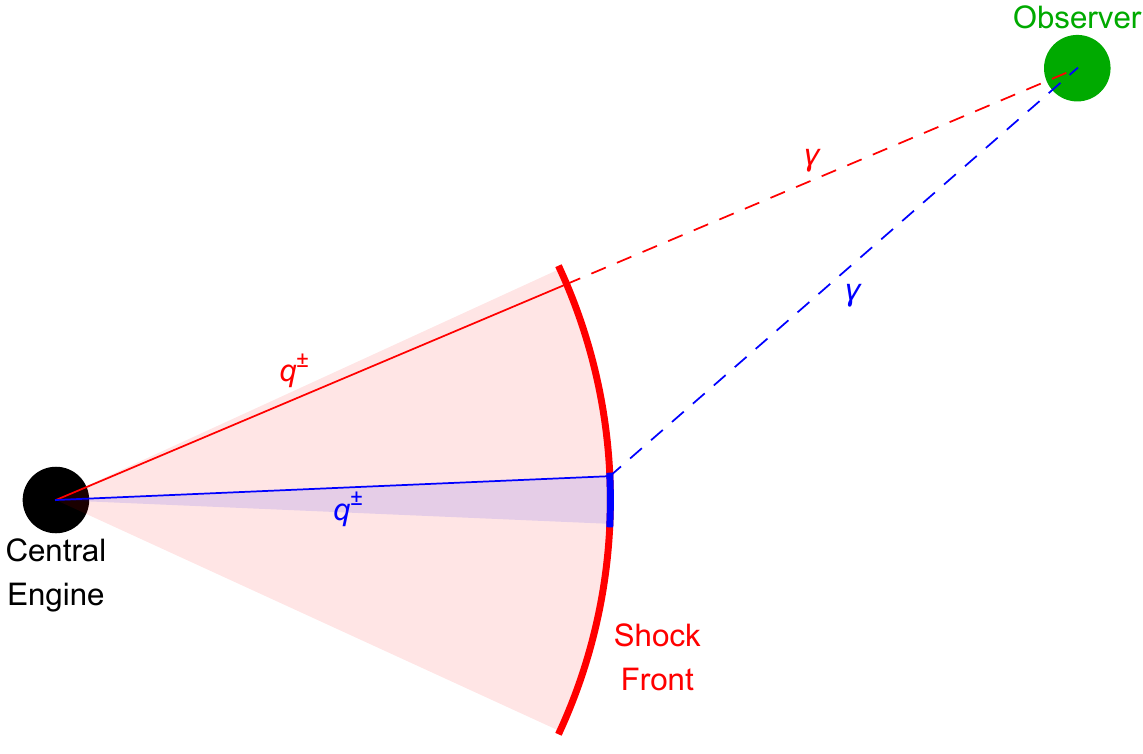}
        \caption{ Model Overview.  The red and blue cones represent
          the regions through which low and high energy charged
          particles propagate.  In the case shown the observer's
          off-axis angle is smaller than the opening angle of the low
          energy jet, so, due to the relativistic beaming effect, the
          most of the observable low energy photons will travel along
          the straight line from the central engine.  Also, the
          observer's off-axis angle is larger that the opening angle
          of the high energy jet, but the high energy radiation will
          still originate near the center of the jet as this is the
          only place where the high energy radiators are located.  The
          observation time of a photon is a sum of two terms. First, the
          time interval spent in plasma as a radiator which
          approximately equals to the distance from the central engine
          to the point of emission. Second, the time interval from
          emission to detection which equals to the distance from the point
          of emission to the observer's location.  For the given position
          of the observer the sum is larger for high energy
          photons.  Therefore, the high energy emission will be
          observed later throughout the burst duration and the
          high energy light curve will be stretched.  }
        \label{fig:modelOverview}
	\end{figure}

	\begin{figure}
		\centering
		\hspace*{\fill}
		\begin{subfigure}{0.45\textwidth}
			\includegraphics[width=\textwidth]{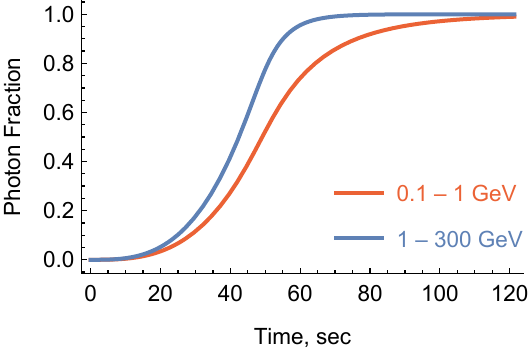}
			\label{fig:sampleLightCurveLogNegative}
			\caption{Stretching factor $\kappa = 0.819$, redshift $z = 1.82$, off-axis angle $\chi = 0$.}
		\end{subfigure}
		\hfill
		\begin{subfigure}{0.45\textwidth}
			\includegraphics[width=\textwidth]{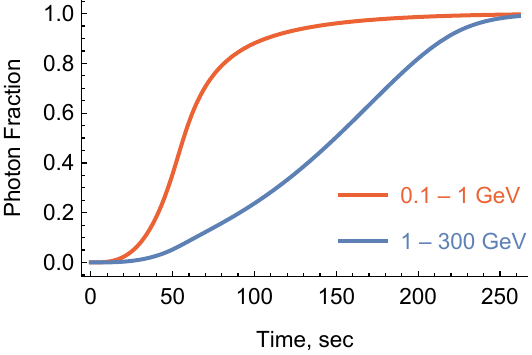}
			\label{fig:sampleLightCurveLogPosivie}
			\caption{Stretching factor $\kappa = 2.43$, redshift $z = 2.106$, off-axis angle $\chi = 5.90 \times 10^{-3}$.}
		\end{subfigure}
		\hspace*{\fill}
		\caption{
			High and low energy light curves produced by the geometrical model.
			Burst parameter values are the same as discussed in Section \ref{sec:parameters}.
		}
		\label{fig:sampleLightCurves}
	\end{figure}

The basic idea of the model is illustrated in Figure
\ref{fig:modelOverview}. The model explains both the stretching
factors lower and higher than 1 with the difference of the position of
the observer with respect to the burst. The sample light curves are
shown in Figure \ref{fig:sampleLightCurves}. The detailed calculations
show that the stretching factors of GRB 090902B and GRB 090926A appear
naturally in the model. The predicted distribution of stretching
factors is shown in Figure \ref{fig:kappaDistribution}.

The rest of the paper is organized as follows. We describe the Fermi
LAT data and the analysis procedures in Section
\ref{sec:observations}. The model and the details on calculations are
explained in Section \ref{sec:model}. The Section \ref{sec:tests}
introduces three phenomenological tests for the model. The parameters
of the model and the new method for estimating the jet observation
angles are shown in Section \ref{sec:results}. The code for modeling
is open source and is publicly available on GitHub
(\url{https://github.com/maxitg/GammaRays}).

\section{Data analysis}
\label{sec:observations}

\subsection{Fermi-LAT Photon Selection}
\label{sec:photonSelection}

\begin{table}
	\centering
	\small
	\begin{tabular}{ l | S[table-format=9.3] | S[table-format=3.3] | S[table-format=-2.3] | S[table-format=1.5] | S[table-format=2.1] | S[table-format=3.1] }
		\multicolumn{1}{ c |}{GRB} & $\mathrm{GBM\ Trigger\ Time}$ & $\mathrm{R.A.}$ & $\mathrm{Dec.}$ & $\mathrm{Location\ Error}$ & $T_{05}$ & $T_{95}$ \\
		\multicolumn{1}{ c |}{Name}& $\mathrm{MET,\ sec}$ & $\mathrm{J2000,\ deg}$ & $\mathrm{J2000,\ deg}$ & $\mathrm{deg}$ & $\mathrm{sec}$ & $\mathrm{sec}$ \\
		\multicolumn{1}{ c |}{\texttt{name}} & $\mathrm{\texttt{time}}$ & \texttt{location.ra} & \texttt{location.dec} & \texttt{location.err} & \texttt{startOffset} & \texttt{endOffset} \\
		\hline
		080825C	&	241366429.105	&	233.9  	&	 -4.5  	&	0.75   	&	 3.2	&	 29.4	\\
		080916C	&	243216766.614	&	119.85 	&	-56.64 	&	0.0001 	&	 5.0	&	209.8	\\
		081006 	&	244996175.173	&	136.32 	&	-62.05 	&	0.52   	&	 0.7	&	115.0	\\
		081024B	&	246576161.864	&	322.95 	&	 21.2  	&	0.22   	&	 0.1	&	191.0	\\
		090217 	&	256539404.560	&	204.83 	&	 -8.42 	&	0.35   	&	 6.2	&	 68.0	\\
		090323 	&	259459364.630	&	190.71 	&	 17.053	&	0.0001 	&	15.9	&	293.9	\\
		090328 	&	259925808.510	&	 90.67 	&	-41.715	&	0.0002 	&	18.8	&	652.9	\\
		090510 	&	263607781.971	&	333.55 	&	-26.583	&	0.0004 	&	 0.6	&	 45.6	\\
		090626 	&	267683530.880	&	170.03 	&	-33.49 	&	0.22   	&	52.2	&	299.9	\\
		090902B	&	273582310.313	&	264.94 	&	 27.324	&	0.001  	&	 7.7	&	825.0	\\
		090926A	&	275631628.990	&	353.4  	&	-66.32 	&	0.01   	&	 5.5	&	225.0	\\
		091003 	&	276237347.585	&	251.52 	&	 36.625	&	0.0005 	&	 3.9	&	452.6	\\
		091031 	&	278683230.850	&	 71.49 	&	-57.65 	&	0.23   	&	 3.1	&	206.2	\\
		100116A	&	285370262.240	&	305.01 	&	 14.43 	&	0.17   	&	 3.0	&	141.0	\\
		100414A	&	292904423.990	&	192.11 	&	  8.693	&	0.0005 	&	17.4	&	288.6	\\
		110120A	&	317231981.230	&	 61.5  	&	-12.0  	&	0.36   	&	 0.5	&	112.8	\\
		110428A	&	325675112.410	&	  5.59 	&	 64.849	&	0.00001	&	10.7	&	407.6	\\
		110721A	&	332916465.760	&	333.2  	&	-38.5  	&	0.20   	&	 0.1	&	239.0	\\
		110731A	&	333803371.954	&	280.504	&	-28.537	&	0.0001 	&	 3.0	&	 24.1
	\end{tabular}
	\caption{Burst data used in our study taken from~\cite{Ackermann:2013zfa}. The third row contains
          the corresponding \texttt{GRBurst} class variable names of
          the software package \url{https://github.com/maxitg/GammaRays/blob/master/GRObservations/bursts}}
	\label{tab:bursts}
\end{table}

First, we download the Pass 7REP (V15) ``SOURCE'' class photons from
the Fermi-LAT data
server~\cite{2009ApJ...697.1071A,Ackermann:2012kna}. In order to
define the regions of interest we use the catalog
\cite{Ackermann:2013zfa}, which contains all the bright gamma-ray
bursts seen by the Fermi LAT since 2008 August 4 to 2011 August 1.
Specifically, we take 4 pieces of information from the tables 2 and 4
of Ref. \cite{Ackermann:2013zfa}:
\begin{itemize}
	\item{
		GBM trigger time.
		It is used as a reference point for the GRB time.
	}
	\item{
		Location of the GRB.
		The burst locations are used to filter out photons
                coming from the other sources.
	}
	\item{
		Location error.
		Location errors are used to improve the accuracy of filtering, specifically to avoid losing statistics by filtering too much.
	}
	\item{
		$T_{05}$ -- $T_{95}$ interval of the LAT-detected
          emission. The beginning of the interval is used as a fixed
          point of stretching. The Fermi-LAT photons are downloaded
          for the period extended in time by $50\%$ towards both past
          and future compared to this interval.
	}
\end{itemize}
The data are summarized in Table \ref{tab:bursts}. For the background
estimation, we also download observational data for 24 hours before
the burst.  The analysis is performed with the Fermi Science Tools
v9r33p0
package~\footnote{\url{http://fermi.gsfc.nasa.gov/ssc/data/analysis/scitools/overview.html}}
following the
guidelines~\footnote{\url{http://fermi.gsfc.nasa.gov/ssc/data/analysis/scitools/data_preparation.html}}.
We require Earth zenith angle to be below $100^\circ$ to remove the
Earth limb emission. After the basic filtering by \texttt{gtselect}
and \texttt{gtmktime} tools is done we perform more elaborate
selection using the point spread function of Fermi LAT, calculated
with \texttt{gtpsf} tool. We use 300 energy bins from \unit[100]{MeV}
to \unit[300]{GeV}, and 300 angular bins with maximal angle being
$30\degree$.  The \texttt{gtltcube} tool is used to compute required
livetime maps. Note, that we calculate PSFs separately for both
conversion types: front and back. We define $\mathrm{PSF}_{95}$ as an
angular distance from the source containing 95\% of the emitted
photons. We require that the photon is closer than $\mathrm{PSF}_{95}
+ \texttt{location.err}$ to the location of the GRB (see Table
\ref{tab:bursts}). The details of the data request and
analysis are given in appendix \ref{sec:fermiCode}.

The set will be used for the analysis with the reservation that it
unavoidably contains some background photons. We consider the
background flux constant in time and estimate it using the data for 24
hours before the burst. In order to calculate the number of background
photons we need an exposure map, which will be discussed in the
following section.

\subsection{Exposure Maps and Background Estimation}

The exposure map is computed with the \texttt{gtexpcube2} tool. We
compute it with 300 energy bins from \unit[100]{MeV} to
\unit[300]{GeV}. The produced \texttt{expcube} files contain exposures
as functions of energy and location on the sky. We use trilinear
interpolation to compute exposures for all energies and
locations. Knowing the exposures, we use the method introduced in the
appendix of \cite{Rubtsov:2011qq} to estimate the background for low
and high-energy bands. Note, that the ``TRANSIENT'' class photons are
not included in the analysis due to the strong dependence of the
background on the spacecraft position.

\subsection{KS-test}

Before applying the KS-test we subtract the background, which is
considered linear for the duration of the burst :

\begin{equation}
	\Phi_i\left(t\right) = \frac{p_i\left(T_1, t\right) - b_i \frac{t-T_1}{T_2-T_1}}{p_i\left(T_1, T_2\right) - b_i}
\end{equation}
Here $\left(T_1, T_2\right)$ is the time range of observations,
$p_i\left(t_1, t_2\right)$ is the number of photons observed in the
time range $t_1$ to $t_2$ in the $i$'s energy band ($i$ is either low
or high for the low and high-energy bands respectively), and $b_i$ is
the estimated number of background photons for the time range $T_1$ to
$T_2$ in the $i$'s energy band.

The number of degrees of freedom used as an input for the KS-test is
$p_i\left(T_1, T_2\right) - b_i$. This is lower than the number of
independent photons and therefore we are conservative when estimating
the probability and significance of the effect. The KS-probabilities
are computed for pairs $\Phi_\text{low}\left(t\right),
\Phi_\text{high}\left(\kappa t\right)$ to obtain the allowed ranges
for stretching factors $\kappa$ for a given confidence level.

\subsection{Stretching factors}

\begin{figure}
        \centering
        \begin{subfigure}{0.49\textwidth}
                \includegraphics[width=\textwidth]{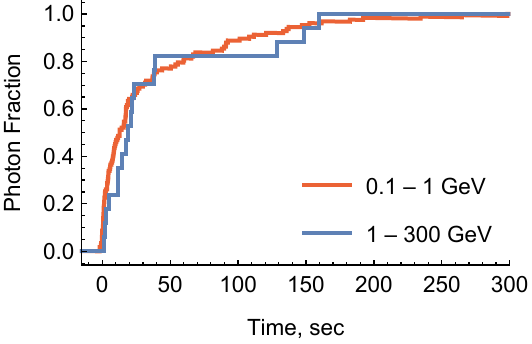}
                \label{fig:lightCurve080916C}
        \end{subfigure}
        \begin{subfigure}{0.49\textwidth}
                \includegraphics[width=\textwidth]{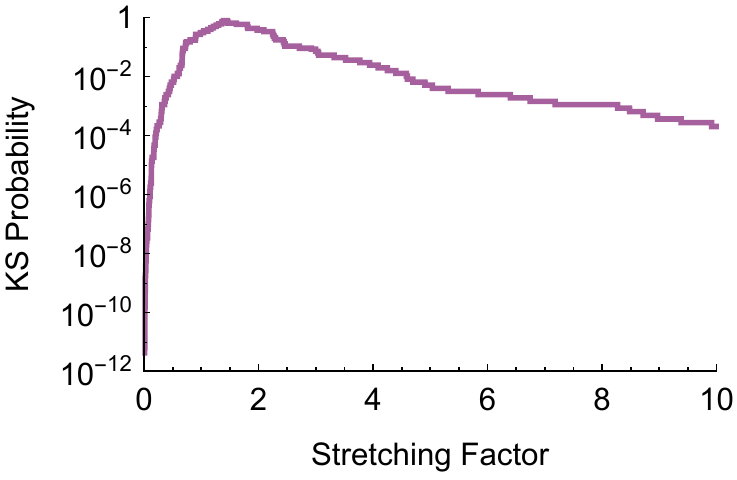}
                \label{fig:probabilities080916C}
        \end{subfigure}
        \caption{GRB 080916C light curves. Stretching factor is compatible with $\kappa = 1$ within $2\sigma$.}
        \label{fig:grb080916C}
\end{figure}

\begin{figure}
        \centering
        \begin{subfigure}{0.49\textwidth}
                \includegraphics[width=\textwidth]{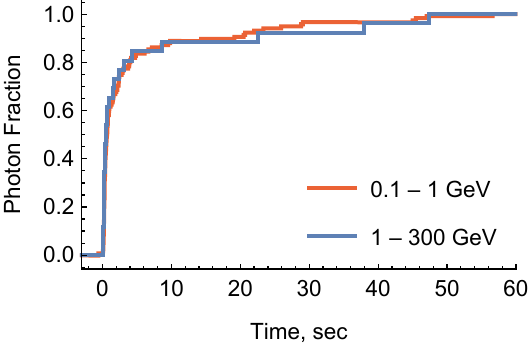}
                \label{fig:lightCurve090510}
        \end{subfigure}
        \begin{subfigure}{0.49\textwidth}
                \includegraphics[width=\textwidth]{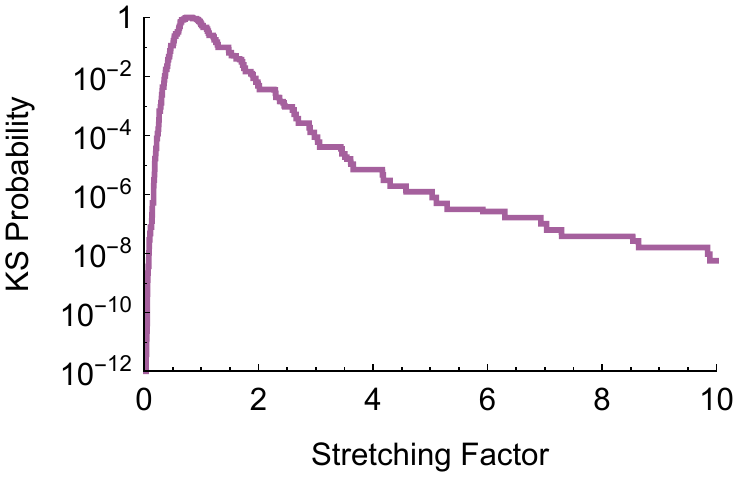}
                \label{fig:probabilities090510}
        \end{subfigure}
        \caption{GRB 090510 light curves. Stretching factor is compatible with $\kappa = 1$ within $1\sigma$.}
        \label{fig:grb090510}
\end{figure}

\begin{figure}
        \centering
        \begin{subfigure}{0.49\textwidth}
                \includegraphics[width=\textwidth]{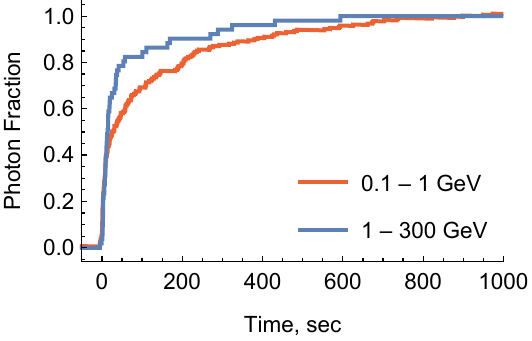}
                \label{fig:lightCurve090902B}
        \end{subfigure}
        \begin{subfigure}{0.49\textwidth}
                \includegraphics[width=\textwidth]{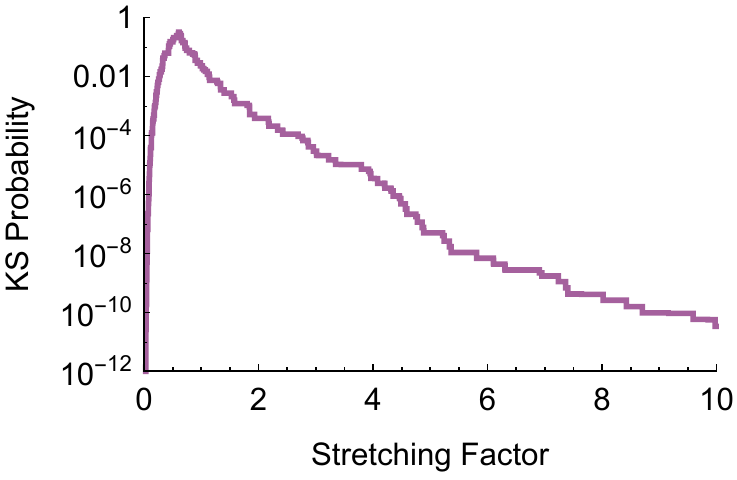}
                \label{fig:probabilities090902B}
        \end{subfigure}
        \caption{GRB 090902B light curves. Low energy radiation is stretched ($\kappa < 1$) with significance of $2.3\sigma$.}
        \label{fig:grb090902B}
\end{figure}

\begin{figure}
        \centering
        \begin{subfigure}{0.49\textwidth}
                \includegraphics[width=\textwidth]{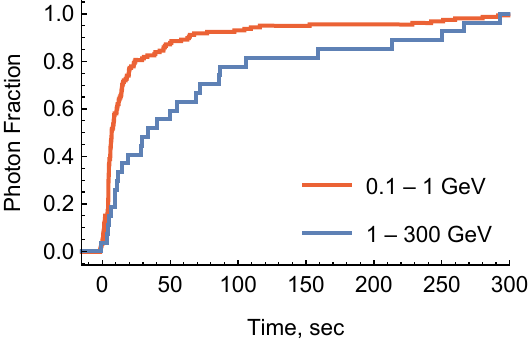}
                \label{fig:lightCurve090926A}
        \end{subfigure}
        \begin{subfigure}{0.49\textwidth}
                \includegraphics[width=\textwidth]{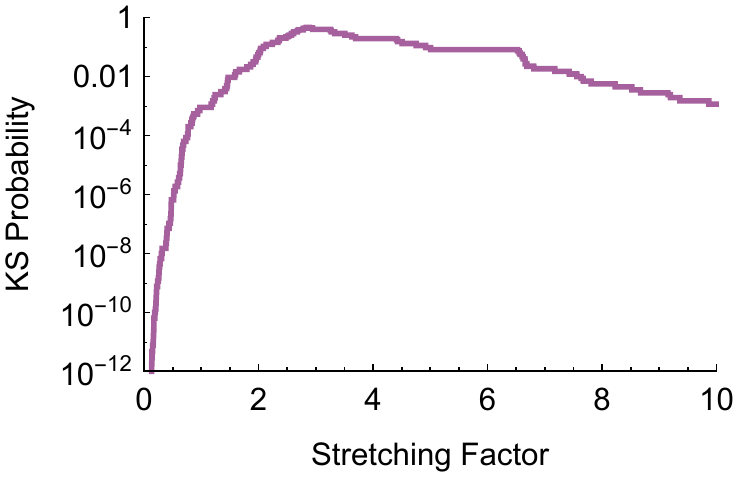}
                \label{fig:probabilities090926A}
        \end{subfigure}
        \caption{GRB 090926A light curves. High energy radiation is stretched ($\kappa > 1$) with significance of $3.3\sigma$.}
        \label{fig:grb090926A}
\end{figure}

Out of 19 bursts studied, only 4 have at least 10 high energy events
remaining after the filtering, and thus eligible for computing the
stretching factors. The results of this computation are shown in
Figures \ref{fig:grb080916C}, \ref{fig:grb090510},
\ref{fig:grb090902B}, \ref{fig:grb090926A} and Table
\ref{tab:observationResults}.

Out of these 4 bursts, two (GRB 080916C and GRB 090510) have
stretching factors compatible with $\kappa = 1$ within $2\sigma$
range.

GRB 090926A has, however, high energy radiation stretched with respect
to low energy radiation (that is $\kappa > 1$). In contrast to that,
GRB 090902B has low energy radiation stretched with respect to high
energy radiation ($\kappa < 1$).

Therefore we obtained an indication that the stretching factors for
observable bursts may take values both larger and smaller than $1$.

\begin{table}
	\centering
	\small
	\begin{tabular}{ l | S[table-format=2.3]@{\; -- \;}S[table-format=2.3] | S[table-format=2.3]@{\; -- \;}S[table-format=2.3] | S[table-format=2.3]@{\; -- \;}S[table-format=2.3] | S[table-format=2.3]@{\; -- \;}S[table-format=2.3] | S[table-format=2.3]@{\; -- \;}S[table-format=2.3] }
		\multicolumn{1}{ c |}{GRB} & \multicolumn{2}{ c |}{$1\sigma$} & \multicolumn{2}{ c |}{$2\sigma$} & \multicolumn{2}{ c |}{$3\sigma$} & \multicolumn{2}{ c |}{$4\sigma$} & \multicolumn{2}{ c }{$5\sigma$} \\
		\hline
		080916C	&	1.04  & 2.24	&	0.67 & 3.32	&	0.42 & 5.83	&	0.19 & 14.9		&	0.087 & 35.7	\\
		090501	&	0.58  & 1.1		&	0.43 & 1.61	&	0.32 & 2.29	&	0.22 &  3.03	&	0.17  &  5.11	\\
		090902B	&	0.604 & 0.613	&	0.35 & 0.89	&	0.22 & 1.53	&	0.13 &  2.86	&	0.078 &  4.49	\\
		090926A &	2.61  & 3.33	&	1.99 & 6.62	&	1.34 & 9.15	&	0.73 & 13.5		&	0.48  & 19.4
	\end{tabular}
	\caption{The ranges of allowed stretching factors for multiple
          levels of significance for the GRBs studied.}
	\label{tab:observationResults}
\end{table}

\section{Model}
\label{sec:model}

The main idea behind our model is to assume that the burst opening
angle depends on the energy of emitted photons, or, equivalently, that the
most energetic plasma particles are concentrated near the axis of the
jet, while low energy particles are on the periphery.  As it's seen 
in Figure \ref{fig:modelOverview} this leads to a time stretching
effect due to relativistic beaming.

Although the details of the exact mechanism of the GRB are not known
there are arguments in support of our model. 

First of all, there are around 750 GRBs detected by GBM, half of which
are in the LAT field of view at the moment of observation
\cite{Vianello:2013ela}.  However, only about 30 of them were detected
by the LAT, and only 4 of them were bright in the high energy band.
If one extrapolates the uniform jet model to the very high energies,
this observation would mean that these groups of bursts are internally
different: some of them produce VHE radiation, while the others do
not.  These differences are hard to explain given that the burst
energetics are similar \cite{Bloom:2003wy}.  These differences in
burst counts find natural explanation in our model, in which the
opening angle of a jet is inversely proportional to the energy of the
photons it radiates.  Therefore the most common scenario is that the
off-axis angle of an observer is smaller than the low energy jet
opening angle, but much larger than the opening angle of a high energy
jet.  Due to that, most of the bursts can only be seen at low
energies.  The 4 bursts we study in this paper were seen, according to
our model, from the lowest off-axis angles.

Another argument comes from consideration of processes happening in
jets, specifically the scattering of particles near the jet boundary.
While plasma particles scatter, they simultaneously lose energy, and
change directions, sometimes propagating beyond the jet boundary, and
therefore increasing the jet opening angle.  So, the processes of
energy loss and increase of jet size are correlated, therefore the low
energy particles should be closer to the jet boundary.

First, let us list the assumptions of the model:

\begin{enumerate}
\item{Time $t = 0$, a spherical shell of plasma is emitted. The center
  is called the central engine.}
\item{The shell points propagate with a constant velocity $v =
  \frac{\sqrt{\gamma^2 - 1}}{\gamma} \sim 1$, so at the time $t$ the
  radius of the shall is $v t$.}
\item{Each point of the shell is an isotropic radiator in its rest
  frame.}
\item{ The radiation intensity is a function of the radiator position
  and the radiation frequency:
	\begin{equation}\label{eq:eta}
		\eta\left(r,\theta,\omega\right) = 
		\frac{\eta_0}{1 + \left(\frac{r}{r_0}\right)^n}
		\exp\left(
			-\left(\frac{\theta}{\theta_0}\right)^2
			\left(\frac{\omega}{\omega_0}\right)^{-2k}
		\right)
		\left(\frac{\omega}{\omega_0}\right)^\alpha
	\end{equation}
	$\eta$ is a number of particles emitted per volume per solid
        angle per frequency. It is a function of the distance $r$ from
        the central engine, of the off-axis angle $\theta$, and of the
        radiation frequency $\omega$.  }
\end{enumerate}

The burst is fully defined by the following set of parameters:
\begin{itemize}
\item{$\gamma$, the relativistic factor of the shell, $\gamma \gg 1$.}
\item{$\eta_0$, which defines the luminosity scale.}
\item{$r_0$, the characteristic jet length; $r_0 \ll \frac{1}{H\left(0\right)}$, $H\left(t\right)$ is the Hubble parameter;}
\item{$n$, which determines the sharpness of the jet end, $n > 3$;}
\item{$\omega_0$, a characteristic radiation frequency;}
\item{$\theta_0$, the opening angle of the jet for radiation with frequency $\omega_0$, $\theta_0 \ll 1$;}
\item{$k$, which determines how much the opening angle changes with frequency, $k < 0$;}
\item{$\alpha$, the bare spectral index, $\alpha < -2k - 1$}
\end{itemize}

In the next step we calculate the observed light curves, and the
stretching factors.

\subsection{Photon observation time}

We begin with computing a time at which some particular photon is
observed. This time is a function of the radiator location $\left(r,
\theta, \phi\right)$, as well as the observer location $\left(d, \chi,
0\right)$ with respect to the burst center (we choose the coordinates
so that the rotation angle of the observer is $0$.) We assume for now
that the observer is too far from the jet to resolve its geometry, yet
close enough so that the expansion of space is negligible. The latter
assumption is made for simplicity of presentation. The cosmological
expansion effect will be restored at the end of this section.

The observation time is a sum of two terms: the time interval from $t
= 0$ to the photon emission (the plasma time), and the time interval
between the emission and the observation (the photon time):
\begin{equation*}
t\left(r, \theta, \phi, d, \chi\right) = t_\text{plasma}\left(r\right) + t_\text{photon}\left(r, \theta, \phi, d, \chi\right)
\end{equation*}
$t_\text{plasma}$ is easy to compute since plasma moves with uniform velocity:
\begin{equation*}
t_\text{plasma}\left(r\right) = \frac{r}{v}
\end{equation*}
$t_\text{photon}$ is a distance between the radiator and the observer:
\begin{align*}
t_\text{photon}\left(r, \theta, \phi, d, \chi\right) &= \sqrt{
	\left( d \cos\chi - r \cos\theta \right)^2 +
	\left( d \sin\chi - r \sin\theta \cos\phi \right)^2 +
	\left( r \sin\theta \sin\phi \right)^2
} \\
&= d\sqrt{
	\left( \cos\chi - \frac{r}{d} \cos\theta \right)^2 +
	\left( \sin\chi - \frac{r}{d} \sin\theta \cos\phi \right)^2 +
	\left(\frac{r}{d} \sin\theta \sin\phi \right)^2
} \\
&\sim d\sqrt{
	\cos^2{\chi} - 2\frac{r}{d} \cos\theta \cos\chi + \sin^2\chi - 2\frac{r}{d} \sin\theta \cos\phi \sin\chi
} \\
&\sim d\left(
	1 - 
	\frac{r}{d} \left( \cos\theta \cos\chi + \sin\theta \cos\phi \sin\chi \right)
\right) \\
&= d - r\left( \cos\theta \cos\chi + \sin\theta \cos\phi \sin\chi \right)
\end{align*}
Combining these expressions together, we get:
\begin{equation*}
t\left(r, \theta, \phi, d, \chi\right) = d + r\left( \frac{1}{v} - \cos\theta \cos\chi - \sin\theta \cos\phi \sin\chi \right)
\end{equation*}
Since no photon can reach the observer faster than the speed of light, and the photons emitted at $t = 0$ reach the observer at $t = d$, this time $t = d$ is the start of the observer's light curve. So we can define a more convenient time origin:
\begin{equation*}
\tau \left(r, \theta, \phi, \chi\right) = t\left(r, \theta, \phi, d, \chi\right) - d = r\left( \frac{1}{v} - \cos\theta \cos\chi - \sin\theta \cos\phi \sin\chi \right)
\end{equation*}
Finally, one can use the spherical law of cosines to write this in terms of the great-circle distance $\sigma\left( \theta, \phi, \chi \right)$ between the points $\left( \theta, \phi \right)$ and $\left( \chi, 0 \right)$:
\begin{equation}
\tau \left(r, \theta, \phi, \chi\right) = r\left( \frac{1}{v} - \cos\sigma\left( \theta, \phi, \chi \right) \right)
\end{equation}

Note that $\tau$ doesn't depend on $d$ anymore. But this is only true if we did not account for cosmology, or if the scale factor did not change from emission to observation. For a distant observer it will change, however, which will stretch the distances between photons:
\begin{equation}
\tau \left(r, \theta, \phi, z, \chi \right) = r\left( \frac{1}{v} - \cos\sigma\left( \theta, \phi, \chi \right) \right) \left( 1 + z \right)
\end{equation}
Here $z$ is the redshift of the burst from the point of view of the observer.

\subsection{Light curve}
We now know enough to compute the observable quantity -- the number of
observed photons $p$. For that we need to integrate the radiation
intensity $\eta$ over the four groups of variables. Two groups are
burst-related: radiator position $\left(r, \theta, \phi\right)$ and
frequency at emission $\omega'$. The other two variables are related
to observer: observation time $\tau$ and frequency at observation
$\omega$. We relate the first two and the second two with the delta
functions:
\begin{align}
p\left( z, \chi; \tau_1, \tau_2; \omega_1, \omega_2 \right) &= \frac{A_\text{det}}{A_\text{ph}\left(z\right)} \int_0^\infty \dif r \int_0^{\pi} r \dif \theta \int_0^{2\pi} r \sin{\theta} \dif \phi \int_0^\infty \dif \omega' \int_{\tau_1}^{\tau_2} \dif \tau \int_{\omega_1}^{\omega_2} \dif \omega \, \eta\left( r, \theta, \omega' \right) \nonumber\\
&\qquad {} \times \underbrace{\frac{1}{\gamma^2\left( 1 - v \cos\sigma\left(\theta, \phi, \chi\right) \right)^2}}_{\text{aberration}} \underbrace{\frac{1}{\gamma\left( 1 - v \cos\sigma\left(\theta, \phi, \chi\right) \right)}}_{\text{time dilation}} \nonumber\\
&\qquad {} \times \delta\left( \frac{\omega'}{\underbrace{\gamma \left( 1 - v\cos\sigma\left(\theta, \phi, \chi\right) \right)}_{\text{relativistic shift}} \underbrace{\left(1+z\right)}_{\text{cosmological shift}}} - \omega \right) \nonumber\\
&\qquad {} \times \delta\left( \tau - r\left( \frac{1}{v} - \cos\sigma\left(\theta, \phi, \chi\right) \right) \left(1+z\right) \right)
\end{align}
Here $A_\text{det}$ is the effective area of the detector, and $A_\text{ph}$ is an area of the sphere over which photons emitted in the burst are spread (see appendix \ref{sec:cosmology} for derivation). We have taken account for four relativistic effects, which affect intensities and frequencies of the radiators: relativistic aberration; time dilation of the radiators relative to the observer; relativistic blue/redshift; and cosmological redshift.

We integrate over $\omega'$ and $r$ using the delta functions. For
that we transform them as follows:
\begin{align*}
\delta\left( \frac{\omega'}{\gamma \left( 1 - v\cos\sigma \right)\left(1+z\right)} - \omega \right) &= \delta\left( \frac{1}{\gamma\left( 1-v\cos\sigma \right)\left(1+z\right)} \left( \omega' - \gamma\left( 1 - v\cos\sigma \right) \left(1+z\right) \omega \right) \right) \\
&= \gamma\left( 1 - v\cos\sigma \right) \left(1+z\right) \delta\left( \omega' - \gamma\left( 1 - v\cos\sigma \right) \left(1+z\right) \omega \right)\\
\delta\left(\tau - r\left( \frac{1}{v} - \cos\sigma \right)\left(1+z\right)\right) &= \delta\left( \left( \frac{1}{v} - \cos\sigma \right) \left( 1+z \right) \left( r - \frac{\tau}{\left(\frac{1}{v} - \cos\sigma\right)\left(1+z\right)} \right) \right) \\
&= \frac{1}{\left(\frac{1}{v} - \cos\sigma\right)\left(1+z\right)} \delta\left( r - \frac{\tau}{\left(\frac{1}{v} - \cos\sigma\right)\left(1+z\right)} \right)
\end{align*}
After the transformations expression for $p$ takes the following form:
\begin{align}
p\left( z, \chi; \tau_1, \tau_2; \omega_1, \omega_2 \right) &= \frac{A_\text{det}}{A_\text{ph}\left(z\right)} \int_0^{\pi} \dif \theta \int_0^{2\pi} \dif \phi \int_{\tau_1}^{\tau_2} \dif \tau \int_{\omega_1}^{\omega_2} \dif \omega \nonumber\\
&\qquad {} \times \eta\left( \frac{\tau}{\left(\frac{1}{v} - \cos\sigma\left(\theta, \phi, \chi\right)\right)\left(1+z\right)}, \theta, \gamma\left( 1 - v\cos\sigma\left(\theta, \phi, \chi\right) \right)\left(1+z\right)\omega  \right) \nonumber\\
&\qquad {} \times \frac{\tau^2 \sin\theta}{\left(\frac{1}{v} - \cos\sigma\left(\theta, \phi, \chi\right)\right)^3 \left(1+z\right)^2 \gamma^2 \left(1-v\cos\sigma\left(\theta, \phi, \chi\right)\right)^2} \nonumber\\
&= \frac{A_\text{det}}{A_\text{ph}\left(z\right)} \frac{1}{v^2 \gamma^2 \left(1+z\right)^2} \int_{\tau_1}^{\tau_2} \dif \tau \, \tau^2 \int_0^{\frac{\pi}{2}} \dif \theta \int_0^{2\pi} \dif \phi \int_{\omega_1}^{\omega_2} \dif \omega \frac{\sin\theta}{\left(\frac{1}{v} - \cos\sigma\left(\theta, \phi, \chi\right)\right)^5} \nonumber\\
&\qquad {} \times \eta\left( \frac{\tau}{\left(\frac{1}{v} - \cos\sigma\left(\theta, \phi, \chi\right)\right)\left(1+z\right)}, \theta, \gamma\left( 1 - v\cos\sigma\left(\theta, \phi, \chi\right) \right)\left(1+z\right)\omega  \right)
\end{align}

Furthermore, one may integrate over $\omega$ and $\tau$ after using an
explicit form for $\eta$ Eq. \ref{eq:eta}:
\begin{align}
p\left( z, \chi; \tau_1, \tau_2; \omega_1, \omega_2 \right) &= \frac{A_\text{det}}{A_\text{ph}\left(z\right)} \frac{1}{v^2 \gamma^2 \left(1+z\right)^2} \int_0^{\pi} \dif \theta \int_0^{2\pi} \dif \phi \int_{\omega_1}^{\omega_2} \dif \omega \int_{\tau_1}^{\tau_2} \dif \tau \, \tau^2 \frac{\sin\theta}{\left(\frac{1}{v} - \cos\sigma\left(\theta, \phi, \chi\right)\right)^5} \nonumber\\
&\qquad {} \times \frac{\eta_0}{1 + \left(\frac{\tau}{r_0\left(\frac{1}{v} - \cos\sigma\left(\theta, \phi, \chi\right)\right)\left(1+z\right)}\right)^n} \nonumber\\
&\qquad {} \times \exp\left(
	-\left(\frac{\theta}{\theta_0}\right)^2
	\left(\frac{\omega\gamma\left( 1 - v\cos\sigma\left(\theta, \phi, \chi\right) \right)\left(1+z\right)}{\omega_0}\right)^{-2k}
\right) \nonumber\\
&\qquad {} \times \left(\frac{\omega\gamma\left( 1 - v\cos\sigma\left(\theta, \phi, \chi\right) \right)\left(1+z\right)}{\omega_0}\right)^\alpha \nonumber\\
&= \frac{A_\text{det}}{A_\text{ph}\left(z\right)} \frac{\eta_0}{\left(v\gamma\left(1+z\right)\right)^{2-\alpha}} \int_0^{\pi} \dif \theta \int_0^{2\pi} \dif \phi \frac{\sin\theta}{\left(\frac{1}{v} - \cos\sigma\left(\theta, \phi, \chi\right)\right)^{5-\alpha}} \nonumber\\
&\qquad {} \times \int_{\omega_1}^{\omega_2} \dif \omega \exp\left(
	-\left(\frac{\theta}{\theta_0}\right)^2
	\left(\frac{\omega\gamma\left( 1 - v\cos\sigma\left(\theta, \phi, \chi\right) \right)\left(1+z\right)}{\omega_0}\right)^{-2k}
\right) \left(\frac{\omega}{\omega_0}\right)^\alpha \nonumber\\
&\qquad {} \times \int_{\tau_1}^{\tau_2} \dif \tau \, \tau^2 \frac{1}{1 + \left(\frac{\tau}{r_0\left(\frac{1}{v} - \cos\sigma\left(\theta, \phi, \chi\right)\right)\left(1+z\right)}\right)^n} \nonumber\\
&= \frac{A_\text{det}}{A_\text{ph}\left(z\right)}
\frac{\eta_0}{\left(v\gamma\left(1+z\right)\right)^{2-\alpha}}
\int_0^{\pi} \dif \theta \int_0^{2\pi} \dif \phi \frac{\sin\theta}{\left(\frac{1}{v} - \cos\sigma\left(\theta, \phi, \chi\right)\right)^{5-\alpha}} \nonumber\\
&\qquad {} \times \left( I\left(z,\chi,\omega_2; \theta,\phi\right) - I\left(z,\chi,\omega_1; \theta,\phi\right) \right) \left( J\left(z,\chi,\tau_2; \theta,\phi\right) - J\left( z,\chi,\tau_1; \theta,\phi \right) \right)
\end{align}
Here $I$ and $J$ are indefinite integrals over $\omega$ and $\tau$:
\begin{align}
I\left(z,\chi,\omega;\theta,\phi\right) &= \frac{
	\omega
	\left(\frac{\omega}{\omega_0}\right)^{\alpha }
	E_{\frac{\alpha +1}{2 k}+1}\left(
		\left(\frac{\theta}{\theta_0}\right)^2
		\left(\frac{\omega}{\omega_0}\right)^{-2k}
		\left(
			\gamma \left(1-v \cos \sigma\left(\theta ,\phi ,\chi \right) \right) \left(1+z\right)
		\right)^{-2 k}
	\right)
}{
	2 k
} \\
J\left(z,\chi,\tau;\theta,\phi\right) &= \frac{\tau^3}{3} \,
_2F_1\left(
	1,
	\frac{3}{n};
	\frac{n+3}{n};
	-\left(\frac{\tau}{
		r_0 \left( \frac{1}{v} - \cos\sigma\left( \theta,\phi,\chi \right) \right) \left(1+z\right)
	}\right)^n
\right)
\end{align}
where exponential integral function $E_n\left(x\right) = \int_1^\infty \frac{e^{-x t} \dif t}{t^n}$.

The remaining integrals over $\theta$ and $\chi$ are hard to do symbolically, so we compute them numerically. To optimize this computation we can use the assumption of small $\theta$ and $\chi$, so that:
\begin{align*}
\sin\theta &\sim \theta \\
\cos\sigma\left(\theta, \phi, \chi \right) &= \cos\theta \cos\chi + \sin\theta \sin\chi \cos\phi \sim 1 - \frac{\theta^2}{2} - \frac{\chi^2}{2} + \theta\chi\cos\phi
\end{align*}
Using this assumption, and the evenness of the integral as a function of $\theta$, we arrive to the optimized expressions for $p$, $I$ and $J$:
\begin{align}
\label{eq:p}
p\left( z, \chi; \tau_1, \tau_2; \omega_1, \omega_2 \right) &= \frac{A_\text{det}}{A_\text{ph}\left(z\right)}
\frac{2\eta_0}{\left(v\gamma\left(1+z\right)\right)^{2-\alpha}}
\int_0^{\infty} \dif \theta \int_0^{\pi} \dif \phi \frac{\theta}{\left(\frac{1}{v} - 1 + \frac{\theta^2}{2} + \frac{\chi^2}{2} - \theta\chi\cos\phi\right)^{5-\alpha}}\nonumber\\
&\qquad {} \times \left( I\left(z,\chi,\omega_2; \theta,\phi\right) - I\left(z,\chi,\omega_1; \theta,\phi\right) \right) \left( J\left(z,\chi,\tau_2; \theta,\phi\right) - J\left( z,\chi,\tau_1; \theta,\phi \right) \right) \\
I\left(z,\chi,\omega;\theta,\phi\right) &= \frac{
	\omega
	\left(\frac{\omega}{\omega_0}\right)^{\alpha }
	E_{\frac{\alpha +1}{2 k}+1}\left(
		\left(\frac{\theta}{\theta_0}\right)^2
		\left(\frac{\omega}{\omega_0}\right)^{-2 k}
		\left(
			v \gamma \left(1+z\right) \left(\frac{1}{v} - 1 + \frac{\theta^2}{2} + \frac{\chi^2}{2} - \theta\chi\cos\phi \right)
		\right)^{-2 k}
	\right)
}{
	2 k
} \\
J\left(z,\chi,\tau;\theta,\phi\right) &= \frac{\tau^3}{3} \,
_2F_1\left(
	1,
	\frac{3}{n};
	\frac{n+3}{n};
	-\left(\frac{\tau}{
		r_0 \left( \frac{1}{v} -  1 + \frac{\theta^2}{2} + \frac{\chi^2}{2} - \theta\chi\cos\phi \right) \left(1+z\right)
	}\right)^n
\right)
\end{align}

Finally, we need to compute the limits of $p$ for $\omega_2
\rightarrow \infty$ and $\tau_2 \rightarrow \infty$. It requires us to
know the limit of $I$ for $\omega \rightarrow \infty$ (here we assume that
$k < 0$):
\begin{equation}
I\left(z,\chi,\infty;\theta,\phi\right) = \lim_{\omega \rightarrow \infty} I\left(z,\chi,\omega;\theta,\phi\right) = 0
\end{equation}
and the limit of $J$ for $\tau \rightarrow \infty$ ($n > 3$ by assumption):
\begin{equation}
J\left(z,\chi,\infty;\theta,\phi\right) = \lim_{\tau \rightarrow \infty} J\left(z,\chi,\tau;\theta,\phi\right) = \left(r_0 \left( \frac{1}{v} -  1 + \frac{\theta^2}{2} + \frac{\chi^2}{2} - \theta\chi\cos\phi \right) \left(1+z\right)\right)^3 \frac{\pi}{n\sin\frac{3\pi}{n}}
\end{equation}

The following quantities of phenomenological importance may be
computed with the use of $p$:
\begin{itemize}
\item{the total number of particles observed in a given energy range,
  $p_\infty\left(z,\chi; \omega_1,\omega_2\right) = p\left( z,\chi;
  0,\infty; \omega_1,\omega_2 \right)$;}
\item{the fraction of photons observed during a given time interval,
  $\Phi\left(z,\chi; \tau_1,\tau_2; \omega_1,\omega_2\right) =
  \frac{p\left( z,\chi; \tau_1,\tau_2;
    \omega_1,\omega_2\right)}{p_\infty\left( z,\chi; \omega_1,\omega_2
    \right)}$;}
\item{the duration of the burst $T_f\left( z,\chi;\omega_1,\omega_2
  \right)$, that is the time by which the fraction $f$ of photons is
  observed. One may compute it by solving the following equation for
  $T_f$: $p\left( z,\chi; 0,T_f; \omega_1,\omega_2\right) = f
  p_\infty\left( z,\chi; \omega_1,\omega_2 \right)$;}
\item{the stretching factor, which will be discussed in the next
  section.}
\end{itemize}

\subsection{Stretching factor}
The stretching factor for a continuous light curve is defined exactly
like the stretching factor for a discrete one. It is the value of
$\kappa$ which makes the KS-distance minimal:
\begin{equation}
\kappa\left(z,\chi; \omega_1, \omega_2, \omega_3\right) = \argmin_\kappa \max_\tau\left| \Phi\left(z,\chi; 0,\tau; \omega_1,\omega_2\right) - \Phi\left( z,\chi; 0,\kappa \tau; \omega_2,\omega_3 \right) \right|
\end{equation}

The maximum of an absolute value is not differentiable, so the
computation of $\kappa$ by the given definition is complicated. One
may instead rewrite this expression in terms of a positive and a
negative KS-distances:
\begin{align*}
D_+\left(z,\chi; \kappa; \omega_1, \omega_2, \omega_3\right) &= \max_\tau\left( \Phi\left(z,\chi; 0,\tau; \omega_1,\omega_2\right) - \Phi\left( z,\chi; 0,\kappa \tau; \omega_2,\omega_3 \right) \right) \\
D_-\left(z,\chi; \kappa; \omega_1, \omega_2, \omega_3\right) &= \min_\tau\left( \Phi\left(z,\chi; 0,\tau; \omega_1,\omega_2\right) - \Phi\left( z,\chi; 0,\kappa \tau; \omega_2,\omega_3 \right) \right) \\
\kappa\left(z,\chi; \omega_1, \omega_2, \omega_3\right) &= \argmin_\kappa \max\left(D_+\left(z,\chi; \kappa; \omega_1, \omega_2, \omega_3\right), -D_-\left(z,\chi; \kappa; \omega_1, \omega_2, \omega_3\right)\right)
\end{align*}

Take note that $\Phi$ monotonously increases with $\tau$. It implies that $D_+$ and $D_-$ monotonously decrease with $\kappa$, so as $D_+ + D_-$. So there is a single value of $\kappa$, for which
\begin{equation}
D_+\left(z,\chi; \kappa; \omega_1, \omega_2, \omega_3\right) = -D_-\left(z,\chi; \kappa; \omega_1, \omega_2, \omega_3\right)
\end{equation}
And this value of $\kappa$ also makes the $\max\left({D_+, -D_-}\right)$ minimal, since $D_+$ monotonously decrease, and $\left(-D_-\right)$ monotonously increase with $\kappa$.

So we arrive at a simpler way to compute $\kappa$ by solving an
equation instead of computing the minimum. 

Being able to compute the stretching factor, we could now compare our
model predictions with observations given the position of an
observer. We don't know the observer's off-axis angle $\chi$, however,
so we cannot check the stretching factor prediction on the
burst-to-burst basis. Instead, we focus on a series of tests, which
will ensure that our model doesn't contradict to existing
observations. Namely the test of the total energy and comparison with
Fermi-LAT data of the distribution of the stretching factors and the
fraction of the high-energy bursts. These tests will be discussed in
the following section.

\section{Observational tests}
\label{sec:tests}
\subsection{Total energy}

In the first test we compute the total energy radiated from the burst,
and ensure that it doesn't exceed the mass of the star from which
the burst originated.

To compute the total energy, we need to multiply the radiation
intensity by frequency, and integrate it over frequencies $\omega$,
volume $\left(r,\theta,\phi\right)$, and observer positions
$\left(\sigma, \xi\right)$. We assume $\theta \ll 1$, and have taken
into account the same relativistic effects as we did in the observed
particle count computation:
\begin{align}
E &= \int_0^\infty \dif \omega \int_0^\infty \dif r \int_0^\infty r \dif \theta \int_0^{2\pi} r \, \theta \dif \phi \int_0^\infty \dif \sigma \int_0^{2\pi} \sin\sigma \dif \xi \nonumber\\
&\quad {} \times \eta\left( r,\theta,\omega \right) \underbrace{\frac{1}{\gamma\left(1-v\cos\sigma \right)}}_{\text{time dilation}} \underbrace{\frac{1}{\gamma^2\left(1-v\cos\sigma \right)^2}}_{\text{aberration}} \underbrace{\frac{\omega}{\gamma\left(1-v\cos\sigma \right)}}_{\text{relativistic shift}} \\
&= \frac{4\pi^2}{\gamma^4} \int_0^\infty \dif \omega \, \omega \int_0^\infty \dif r \, r^2 \int_0^\infty \dif \theta \, \theta \, \eta\left( r,\theta,\omega \right) \int_0^\infty \dif \sigma \frac{\sin\sigma}{\left(1-v\cos\sigma \right)^4} \nonumber
\end{align}
An integral over $\sigma$ is computable analytically:
\begin{equation*}
\int_0^\infty \dif \sigma \frac{\sin\sigma}{\left(1-v\cos\sigma \right)^4} = \frac{2\left(3+v^2\right)}{3\left(1-v^2\right)^3} = \frac{2}{3}\gamma^6 \left(4 - \frac{1}{\gamma^2}\right)
\end{equation*}
Substituting this result into the expression for $E$ we get:
\begin{equation}
E = \frac{32\pi^2}{3}\left(\gamma^2 - \frac{1}{4}\right) \int_0^\infty \dif \omega \, \omega \int_0^\infty \dif r \, r^2 \int_0^\infty \dif \theta \, \theta \, \eta\left( r,\theta,\omega \right)
\end{equation}
Now we can use the expression for $\eta$ to do the remaining integrals:
\begin{align*}
E = \frac{32\pi^2 \eta_0}{3}\left(\gamma^2 - \frac{1}{4}\right) \int_0^\infty \dif \omega \, \omega \left(\frac{\omega}{\omega_0}\right)^\alpha \int_0^\infty \dif r \, \frac{r^2}{1 + \left(\frac{r}{r_0}\right)^n} \int_0^\infty \dif \theta \, \theta \, \exp\left(-\left(\frac{\theta}{\theta_0}\right)^2\left(\frac{\omega}{\omega_0}\right)^{-2k}\right)
\end{align*}
The integrals over $r$ and $\theta$ can be computed symbolically:
\begin{equation*}
\int_0^\infty \dif r \, \frac{r^2}{1 + \left(\frac{r}{r_0}\right)^n} = \frac{\pi}{n\sin\left(\frac{3\pi}{n}\right)}r_0^3
\end{equation*}
\begin{equation*}
\int_0^\infty \dif \theta \, \theta \, \exp\left(-\left(\frac{\theta}{\theta_0}\right)^2\left(\frac{\omega}{\omega_0}\right)^{-2k}\right) = \frac{1}{2}\theta_0^2 \left(\frac{\omega}{\omega_0}\right)^{2k}
\end{equation*}
Now we have only one integral left:
\begin{equation}
E = \frac{16\pi^3}{3 n\sin\left(\frac{3\pi}{n}\right)} \left(\gamma^2 - \frac{1}{4}\right) \, \eta_0 \, r_0^3 \, \theta_0^2 \int_0^\infty \dif \omega \, \omega \left(\frac{\omega}{\omega_0}\right)^{2k+\alpha}
\end{equation}

Note, however, that since $\alpha < -2k-1$ this integral diverges for
$\omega \rightarrow 0$. This is not a problem as long as our model is
not expected to describe low-energy radiation of the burst. One should
ensure that the total energy of the high-energy radiation is not too
large. For this test, we integrate only over those radiators which
emission is observed by Fermi LAT, that is over the frequencies
greater than $\frac{\omega_1}{\gamma}$, where $\omega_1$ is the
smallest observable photon frequency. Now we compute the last integral
and get the final expression for $E$:
\begin{align}
E\left(\omega_1\right)
&= \frac{16\pi^3}{3 n\sin\left(\frac{3\pi}{n}\right)} \left(\gamma^2 - \frac{1}{4}\right) \, \eta_0 \, r_0^3 \, \theta_0^2 \int_{\frac{\omega_1}{\gamma}}^\infty \dif \omega \, \omega \left(\frac{\omega}{\omega_0}\right)^{2k+\alpha} \nonumber\\
&= \frac{16\pi^3}{3 n\sin\left(\frac{3\pi}{n}\right)} \left(\gamma^2 - \frac{1}{4}\right) \, \eta_0 \, r_0^3 \, \theta_0^2 \frac{\omega_1^2 \left(\frac{\omega_1}{\gamma \omega_0}\right)^{2k+\alpha}}{\gamma ^2 (-2k-\alpha-2)} \nonumber\\
&= \frac{16\pi^3}{3 n\left(-2k-\alpha-2\right)\sin\left(\frac{3\pi}{n}\right)} \gamma^{-2k-\alpha} \left(1 - \frac{1}{4\gamma^2}\right) \, \eta_0 \, r_0^3 \, \theta_0^2 \frac{\omega_0^{-2k-\alpha}}{\omega_1^{-2k-\alpha-2}}
\end{align}

This energy should not exceed a typical mass $M_s$ of a massive
star. So, finally, we arrive at the first constraint for the burst
parameters:
\begin{equation}
\frac{16\pi^3}{3 n\left(-2k-\alpha-2\right)\sin\left(\frac{3\pi}{n}\right)} \gamma^{-2k-\alpha} \left(1 - \frac{1}{4\gamma^2}\right) \, \eta_0 \, r_0^3 \, \theta_0^2 \frac{\omega_0^{-2k-\alpha}}{\omega_1^{-2k-\alpha-2}} < M_s
\end{equation}

\subsection{Distribution of stretching factors}
\label{sec:distribution}

The second test is to calculate the distribution of stretching factors
of observable bursts, and to compare it to observations.

The computation of the exact and precise distribution is technically
hard and computationally intensive because stretching factor is not a
monotonic function of neither redshift nor off-axis angles.  We do not
compute the distribution directly. Instead, we use the Monte-Carlo
method to produce a large representative sample of stretching
factors. Then, we calculate an empirical CDF of this sample, which can
be compared with observations using the KS-test. Since one may compute
much larger sample than that of observations, there is no precision
loss due to this simplification.

We still have one ingredient missing though -- the evolution of the
bursts density. We assume that the bursts density is roughly
proportional to the stars density, and the latter is roughly
proportional to the matter density, which changes with $z$ as
$\left(1+z\right)^3$. It is clear, however, that since no stars
existed at very small redshifts, the burst density should decline
there, so we add an exponential cutoff to it:
\begin{equation}\label{eq:redshift}
\rho = \rho_0 \left(1+z\right)^3 \exp\left(-\frac{z}{z_c}\right)\,
\end{equation}
where $\rho_0$ is a normalization factor, and $z_c$ is a redshift
scale, after which the density is cut off.

For the modeling we first need to define the range of redshifts and
off-axis angles. The range should include all the observable jet
positions, but one should keep the range as small as possible in order
to avoid production of too many points corresponding to invisible
bursts. For simplicity we consider the rectangular region.

We start with selecting a range for redshifts. Ideally, one would
expect that this range starts with $z=0$. Then, however, there will be
visible jets for all possible off-axis angles, which will make our
angles range too large. Note also, that since the bursts count
increases with redshift as $z^3$, the probability to observe a
low-redshift burst is very small. So instead of selecting the smallest
redshift to be $0$, we select it to be a small number $z_\text{min} = 0.1$.

$z_\text{max}$ is defined by the farthest jet, which can be
observed. Since the burst observability
$p_\infty\left(z,\chi;\omega_2,\omega_3\right)$ decreases with both
redshift and $\chi$, we can find the maximum redshift by solving the
following equation:
\begin{equation}
p_\infty\left(z_\text{max},0;\omega_2,\omega_3\right) = p_\text{min} \,,
\end{equation}
where $p_\text{min}$ is the minimum number of particles required to claim an observation.

The observability declines with $z$ and $\chi$, so the observable
burst with the largest $\chi$ should be located at the redshift
$z_\text{min}$. We can find this angle by solving the similar
equation, as we did for the redshifts:
\begin{equation}
  	p_\infty\left(z_\text{min},\chi_\text{max};\omega_2,\omega_3\right) = p_\text{min}
\end{equation}

The simulation is performed by repetition of the following 3 steps
before the required number of bursts is attained:
\begin{enumerate}
\item{
First, the random redshift is generated following the distribution
\ref{eq:redshift}. This may be done with the Inverse transform sampling
method. The CDF of the distribution of redshifts is a ratio of
 redshift counts in different space volumes:
	\begin{equation}
	\Phi_z\left(z\right) = \frac{\int_{z_\text{min}}^z \rho\left(z'\right) \dif V\left(z'\right)}{\int_{z_\text{max}}^{z_\text{max}} \rho\left(z'\right) \dif V\left(z'\right)}
	\end{equation}
	where $\dif V\left(z'\right)$ is the volume of the
        infinitesimal shell surrounding a sphere over which bursts at
        redshift $z'$ are distributed (see appendix
        \ref{sec:cosmology} for derivation).

To generate a redshift, one should uniformly select a value of
$\Phi_z$, and solve the corresponding equation for $z$:
	\begin{equation}
	\Phi_z\left(z\right) = x
	\end{equation}
where $x$ is a random variable uniformly distributed in the range $0$ to $1$.
}
\item{ The random off-axis angle may be then generated with a similar
  method. The CDF of the angles distribution is a ratio of spherical
  areas:
	\begin{equation}
	\Phi_\chi\left(\chi\right) = \frac{\int_0^\chi \sin\chi' \dif\chi'}{\int_0^{\chi_\text{max}} \sin\chi' \dif\chi'} \approx \frac{\int_0^\chi \chi' \dif\chi'}{\int_0^{\chi_\text{max}} \chi' \dif\chi'} = \left(\frac{\chi}{\chi_\text{max}}\right)^2\,.
	\end{equation}

	As with redshifts, we get a properly distributed $\chi$ by solving the equation:
	\begin{align}
	\Phi_\chi\left(\chi\right) = \left(\frac{\chi}{\chi_\text{max}}\right)^2 = y\,, \\
	\chi = \chi_\text{max}\sqrt{y}\,,
	\end{align}
	where $y$ is an another independent random variable uniformly distributed in the range $0$ to $1$.
}
\item{Now we check, whether the burst in a selected position may be
  observed: $p_\infty\left(z,\chi;\omega_2,\omega_3\right) >
  p_\text{min}$. If it is, add $\kappa\left(z,\chi\right)$ to the
  sample.}
\end{enumerate}

With this algorithm, we arrive to the stretching factor distribution
which may be compared with the observed one. 

\subsection{High energy bursts fraction}

Our final test is based on the answer to the following question: given
the bursts observed in the whole energy range
$\left(\omega_1,\omega_3\right)$, which fraction of them may also be
observed in the high energy range $\left(\omega_2,\omega_3\right)$?

To calculate this ratio, one needs to compute the number of bursts
visible in a given energy range, which is the integral over space and
jet directions:
\begin{align}
b\left(\omega_1,\omega_2\right) &= \int_0^{z_\text{max}\left(\omega_1,\omega_2\right)} \dif V\left(z\right) \rho\left(z\right) \int_0^{\chi_\text{max}\left(z,\omega_1,\omega_2\right)} 2\pi \sin\chi \dif\chi \nonumber\\
&\approx 2\pi \int_0^{z_\text{max}\left(\omega_1,\omega_2\right)} \dif V\left(z\right) \rho\left(z\right) \int_0^{\chi_\text{max}\left(z,\omega_1,\omega_2\right)} \chi \dif\chi = \pi \int_0^{z_\text{max}\left(\omega_1,\omega_2\right)} \dif V\left(z\right) \rho\left(z\right) \chi^2_\text{max}\left(z,\omega_1,\omega_2\right)\,,
\end{align}
where $z_\text{max}\left(\omega_1,\omega_2\right)$ and
$\chi_\text{max}\left(z,\omega_1,\omega_2\right)$ are the same values
we used in the previous section: the maximum redshift from which a
burst can be observed in a given energy range, and the maximum
off-axis angle with which the burst at redshift $z$ can be observed.

The required fraction is the ratio of these integrals:
\begin{equation}
f\left(\omega_1,\omega_2,\omega_3\right) = \frac{b\left(\omega_2,\omega_3\right)}{b\left(\omega_1,\omega_3\right)} = \frac{\int_0^{z_\text{max}\left(\omega_2,\omega_3\right)} \dif V\left(z\right) \rho\left(z\right) \chi^2_\text{max}\left(z,\omega_2,\omega_3\right)}{\int_0^{z_\text{max}\left(\omega_1,\omega_3\right)} \dif V\left(z\right) \rho\left(z\right) \chi^2_\text{max}\left(z,\omega_1,\omega_3\right)}\,.
\end{equation}

Our 3rd test is the comparison of the obtained ratio with the observed value.

\section{Results}
\label{sec:results}

\subsection{Parameter fit}
\label{sec:parameters}

For now we have discussed how to compute various observables of the model,
and how to test the model against the data from observations. Now we
come to the discussion of specific values for the parameters of the
burst.

First of all, we need to determine, which parameters to fit. At most,
the observables of a particular burst depend on 10 parameters: 8 of
the burst $\left(\gamma, \eta_0, r_0, n, \omega_0, \theta_0, k,
\alpha\right)$, and 2 of the observer's position related to the burst
$\left(z, \chi\right)$. However, some of these parameters only change
observables trivially, and, also, there are transformations of
variables, which do not change observables at all (see equation
\ref{eq:p}):
\begin{itemize}
	\item{The number of photons $p$ depends linearly on $\eta_0$,
          so $\eta_0$ doesn't affect other observables like duration,
          or stretching factor. One can easily fit $\eta_0$ to match
          the observed photon count.}
	\item{If we make a simultaneous transformation of $r_0
          \rightarrow \lambda r_0$ and $\eta_0 \rightarrow
          \frac{1}{\lambda^3}\eta_0$, where $\lambda$ is a parameter
          of transformation, only duration of the burst will
          change. Other observables, like the total number of photons,
          or stretching factor, will not be affected.}
	\item{Finally, if we make a following transformation:
          $\omega_0 \rightarrow \lambda \omega_0$, $\theta_0
          \rightarrow \lambda^k \theta_0$, $\eta_0 \rightarrow
          \lambda^\alpha \eta_0$, where $\lambda$ is another
          transformation parameter, no observables will change
          whatsoever.}
\end{itemize}
Given these transformations, one can reduce the number of parameters
for fit to 7: $\left(\gamma, n, \theta_0, k, \alpha, z, \chi\right)$.

Also, some of these parameters are known from observations:
\begin{itemize}
	\item{Redshifts $z$ of many bursts have been measured
          \cite{Ackermann:2013zfa}.}
	\item{We know that relativistic factors $\gamma$ are on the
          order of magnitude\footnote{Our minimization procedure can
            find similar minimums for other values of $\gamma$, at
            least from $100$ to $1000$} of $\gamma = 300$
          \cite{Ghirlanda:2011bn}.}
\end{itemize}
This knowledge allows us to reduce the number of parameters to minimize against to just 5: $\left(n, \theta_0, k, \alpha, \chi\right)$.

The best fit parameters are found by minimizing the cost function
$C\left(n, \theta_0, k, \alpha, \chi\right)$ which should
satisfy the following objectives:
\begin{itemize}
	\item { Total energy of the burst is finite (that is
          $k+\frac{\alpha}{2}+1 < 0$) and smaller than $\unit[6 \times
            10^{53}]{GeV} \approx \unit[10^{51}]{erg}$
          \cite{Gehrels:2013xd}.  }
	\item { The stretching factor of the burst should be
          compatible with the value for the GRB 090926A.  }
	\item { The ratio of photons with high energies and low
          energies should be compatible with the value for the GRB
          090926A, $\frac{p_{\infty}
            \left(z,\chi;\unit[1]{GeV},\infty\right)}{p_{\infty}
            \left(z,\chi;\unit[0.1]{GeV},\unit[1]{GeV}\right)} =
          0.057$.  }
	\item { The burst should not be too faint compared to
          other bursts from the sample. In other words, the total
          number of observed photons should approximately equal to the
          median among the sample.  }
	\item { We assume that all bursts have the same burst
          parameters $\left(\gamma, \eta_0, r_0, n, \omega_0,
          \theta_0, k, \alpha\right)$ and require that bursts with
          small stretching factors like GRB 090902B should be possible
          to observe. So we require that bursts with the stretching
          factor of GRB 090902B or lower appear in random samples by
          varying $z$ and $\chi$ (sampling is done the same way as in
          Section \ref{sec:distribution}). Note that this requirement
          unlike others applies to an ensemble of simulated bursts. }
\end{itemize}

The cost function is then computed by the following procedure:
\begin{enumerate}
	\item{
		Set $\gamma = 300$, $\omega_0 = \unit[1]{GeV}$, $z = 2.1062$ (which is the redshift of GRB 090926A)
	}
	\item{
		Set $\eta_0$ and $r_0$ such that the duration of the burst, and the total number of observed photons are compatible with GRB 090926A, $T_{0.99}\left(z,\chi;\unit[0.1]{GeV},\infty\right) = \unit[219.5]{sec}$ and $p_{\infty} \left(z,\chi;\unit[0.1]{GeV},\infty\right) = 179.996$
	}
	\item{
		If $k+\frac{\alpha}{2}+1 < 0$, or the energy of the burst $E\left(\unit[0.1]{GeV}\right) > \unit[6 \times 10^{53}]{GeV}$, the cost function equals to the penalization factor: $C\left(n, \theta_0, k, \alpha, \chi\right) = 400$.
	}
	\item{
		\label{item:sample}
		Compute the small sample of 10 bursts (we will need stretching factors and total photon counts) with the same fixed burst parameters $\left(\gamma, \eta_0, r_0, n, \omega_0, \theta_0, k, \alpha\right)$, and with $z$ and $\chi$ representatively distributed (as discussed in Section \ref{sec:distribution}).
	}
	\item{
		Compute the cost due to the stretching factor:
		\begin{equation}
			C_\kappa = \frac{
				\log{\kappa\left(z,\chi;\unit[0.1]{GeV},\unit[1]{GeV},\infty\right)} - \log{\kappa_{\text{GRB090926A}}}
			}{\Delta\log\kappa_{\text{GRB090926A}}}
		\end{equation}
		Here $\log\kappa_{\text{GRB090926A}} = \frac{1}{2}\left(\log\left(6.62\right) + \log\left(1.99\right)\right)$ and $\Delta\log\kappa_{\text{GRB090926A}} = \frac{1}{2}\left(\log\left(6.62\right) - \log\left(1.99\right)\right)$.
	}
	\item{
		Compute the cost due to the fraction of high and low energy photon counts:
		\begin{equation}
			C_f = \frac{
				\log{\frac{
					p_{\infty} \left(z,\chi;\unit[1]{GeV},\infty\right)
				}{
					p_{\infty} \left(z,\chi;\unit[0.1]{GeV},\unit[1]{GeV}\right)}
				} - \log f_{\text{GRB090926A}}
			}{\log(10)}
		\end{equation}
		Here $f_{\text{GRB090926A}} = 0.0570003$.
	}
	\item{
		Compute the cost due to the brightness of the burst compared to the median:
		\begin{equation}
			C_b = \frac{
				\log{\frac{
					p_{\infty} \left(z,\chi;\unit[0.1]{GeV},\infty\right)
				}{
					p_\text{med}
				}} - 0
			}{\log(10)}
		\end{equation}
		Here $p_\text{med}$ is the median number of observed photons among the sample computed in the step \ref{item:sample}.
	}
	\item{
		Compute the cost due to the minimal stretching factor from the sample:
		\begin{equation}
			C_{\kappa\text{min}} = \max\left(0, 
				\frac{
					\log{\kappa_\text{min}} - \log{\kappa_{\text{GRB090902B}}}
				}{\Delta\log\kappa_{\text{GRB090902B}}}\right)
		\end{equation}
		Here $\kappa_\text{min}$ is the minimal stretching factor from the sample computed in the step \ref{item:sample}, $\log\kappa_{\text{GRB090902B}} = \frac{1}{2}\left(\log\left(0.89\right) + \log\left(0.35\right)\right)$ and $\Delta\log\kappa_{\text{GRB090902B}} = \frac{1}{2}\left(\log\left(0.89\right) - \log\left(0.35\right)\right)$.
	}
	\item{
		Finally, the value of the cost function is the sum of squares of the four:
		\begin{equation}
			C = C_\kappa^2 + C_f^2 + C_b^2 + C_{\kappa\text{min}}^2
		\end{equation}
	}
\end{enumerate}

\begin{table}
	\centering
	\small
	\begin{tabular}{ S[table-format=2.4] | S[table-format=1.5] | S[table-format=-1.6] | S[table-format=-1.7] | S[table-format=1.8] }
		$\mathrm{n}$ & $\mathrm{\theta_0}$ & $\mathrm{k}$ & $\mathrm{\alpha}$ & $\mathrm{\chi}$ \\
		\hline
		22.1525	&	$\mathrm{2.16424 \times 10^{-4}}$ 	&	-0.417021  	&	-1.3516300 	&	$\mathrm{7.04982 \times 10^{-3}}$	\\
		25.8640	&	$\mathrm{4.90222 \times 10^{-8}}$	&	-2.908750 	&	 3.5380100 	&	$\mathrm{2.86570 \times 10^{-4}}$ 	\\
		17.5380	&	$\mathrm{1.46284 \times 10^{-3}}$ 	&	-2.133600 	&	-2.1336000 	&	$\mathrm{1.52624 \times 10^{-3}}$ 	\\
		17.9736	&	$\mathrm{4.52631 \times 10^{-6}}$	&	-1.402620 	&	-0.0114132 	&	$\mathrm{3.25900 \times 10^{-4}}$ 	\\
		 5.0000	&	$\mathrm{1.12535 \times 10^{-7}}$	&	-0.200000	&	-2.0000000 	&	$\mathrm{4.73795 \times 10^{-3}}$ 	\\
		 7.0000	&	$\mathrm{2.00000 \times 10^{-12}}$	&	-3.000000	&	-3.0000000	&	$\mathrm{1.73795 \times 10^{-3}}$
	\end{tabular}
	\caption{Initial points used for minimization procedure from Section \ref{sec:parameters}.}
	\label{tab:fitInitialPoints}
\end{table}

We use the Nelder-Mead (downhill simplex) method for the minimization
procedure. Initial points were chosen to include the areas in
parameter space where each cost is close to $0$ (first $4$ points
correspondingly), and to cover the large fraction of the parameter
space. You can see the values in Table \ref{tab:fitInitialPoints}.

The minimization converged to the following parameter values:
\begin{multicols}{2}
\begin{itemize}
		\item{$\gamma = 300$}
		\item{$\eta_0 = \unit[4.629640817921788 \times 10^{34}]{sec^{-3} GeV^{-1}}$}
		\item{$r_0 = \unit[3.142715108207946 \times 10^6]{sec}$}
		\item{$\bm{n = 22.9977}$}
		\item{$\omega_0 = \unit[1]{GeV}$}
		\item{$\bm{\theta_0 = 7.93393 \times 10^{-5}}$}
		\item{$\bm{k = -0.65356}$}
		\item{$\bm{\alpha = -0.724394}$}
		\item{$z = 2.1062$}
		\item{$\bm{\chi = 0.00590163}$}
\end{itemize}
\end{multicols}

Note, that these parameter values (except for $z$ and $\chi$) are
universal, and they may describe the whole population of observed
bursts, as will be shown in the next section.

\subsection{Results of the tests}
	
As was discussed in the previous section our model can reproduce both
stretching factors smaller than $1$ (like of GRB 090902B) and larger
than $1$ (like of GRB 090926A) depending on redshift and observer's
off-axis angle (see fig. \ref{fig:sampleLightCurves}).

\begin{figure}
	\hspace*{\fill}
	\begin{subfigure}{0.45\textwidth}
		\includegraphics[width=\textwidth]{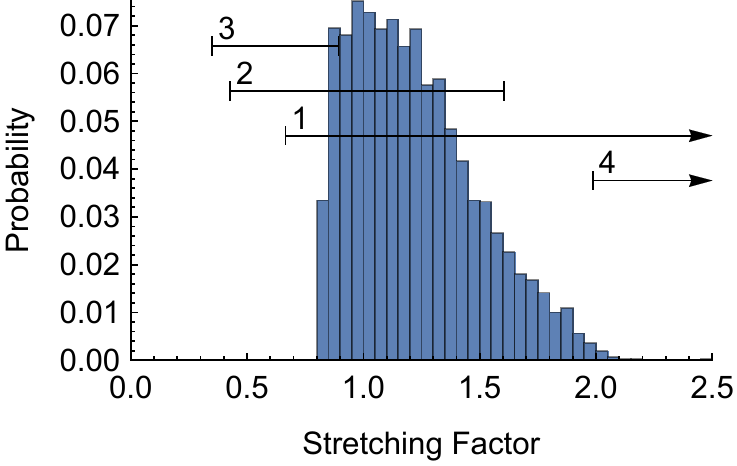}
		\label{fig:kappaDistributionHistogram}
		\caption{Histogram (blue) obtained from the model. Black error bars show observed stretching factors.}
	\end{subfigure}
	\hfill
	\begin{subfigure}{0.45\textwidth}
		\includegraphics[width=\textwidth]{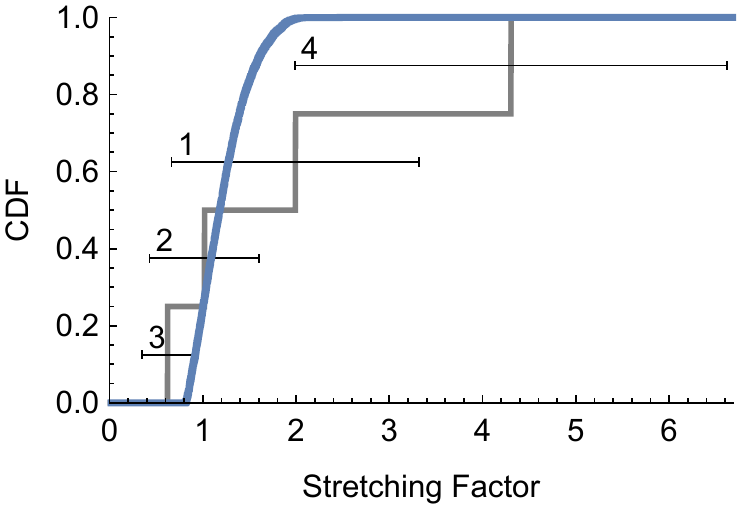}
		\label{fig:kappaDistributionCDF}
		\caption{CDF (blue) obtained from the model. Gray curve and black error bars show observed stretching factors.}
	\end{subfigure}
	\hspace*{\fill}
	\caption{
		Stretching factors histogram and CDF produced by our model.
		The sample contains $4096$ bursts.
		Numbered error bars correspond to GRB 080916C, GRB 090510, GRB 090902B and GRB 090926A (in this order).
	}
	\label{fig:kappaDistribution}
\end{figure}

Below are the results of the tests of Section~\ref{sec:tests}:
\begin{itemize}
	\item{ The total energy emitted in $\unit[100]{MeV}$ and more
          energetic gamma rays $E < \unit[5.89 \times 10^{53}]{GeV}$,
          which is in agreement with \cite{Gehrels:2013xd}.  }
	\item{ The distribution of observable bursts' stretching
          factors is shown in Fig. \ref{fig:kappaDistribution}.  This
          distribution does not contradict to the values of stretching
          factors obtained in Section \ref{sec:observations}.  }
	\item{ The fraction of bursts observable in low energy band,
          which are also observed in high energy band is $f_m = 0.110$.
          The Fermi-LAT catalog contains $35$ bursts out of which $4$
          can be observed in high energy band.  Therefore the observed
          value $f_o = 0.11 \pm 0.05$ (error from binomial
          distribution) agrees with the model.  }
\end{itemize}

Therefore the model successfully explains the variety of the
stretching factors observed by the Fermi LAT.

\section{Discussion}

\begin{figure}
        \centering
        \includegraphics[width=1.0\textwidth]{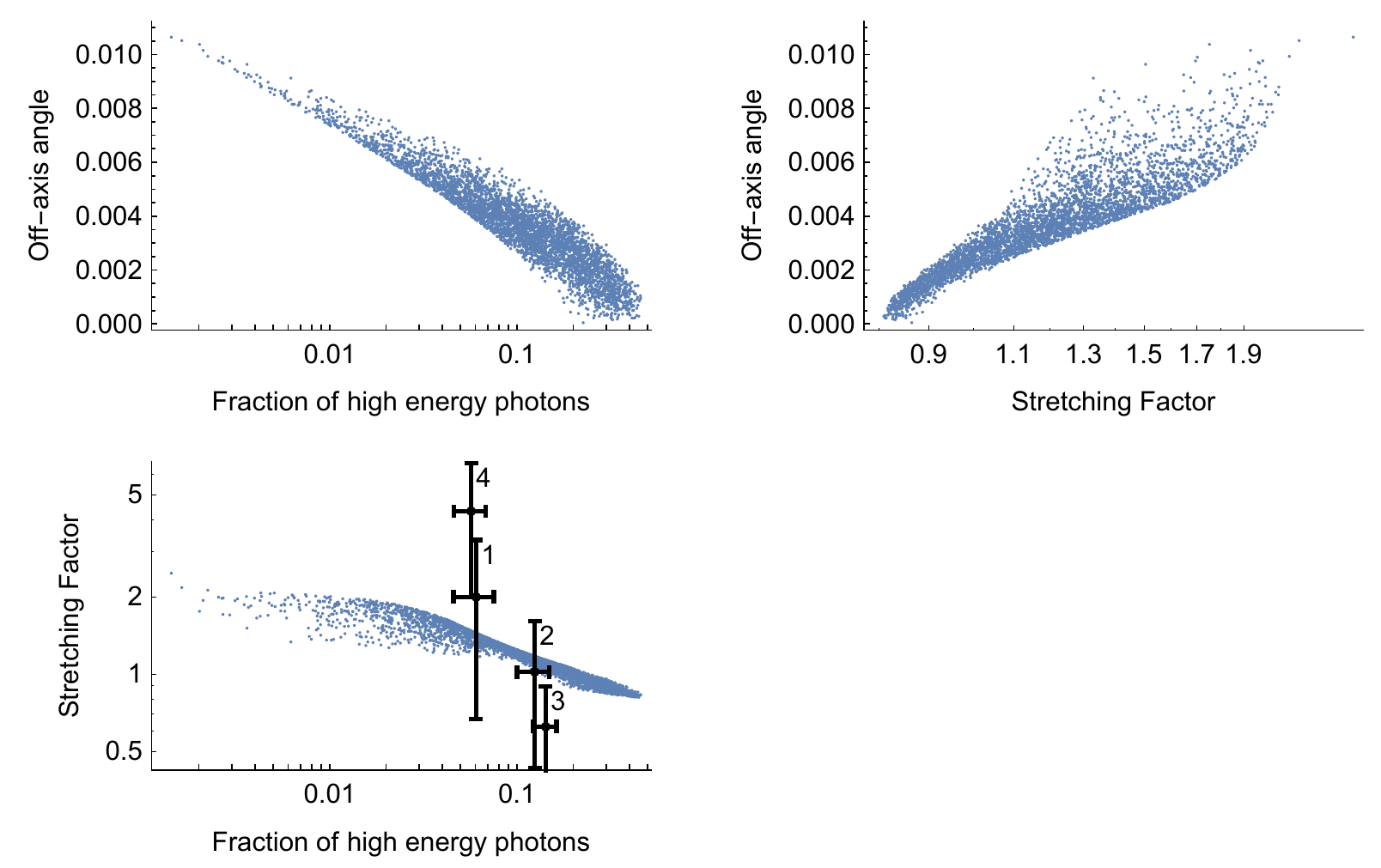}
        \caption{
        	Correlations between off-axis angles, stretching factors and high to low energy photon count ratios found in the sample produced by our model.
        	The sample contains $4096$ bursts.
        	Numbered black crosses show data for GRB 080916C, GRB 090510, GRB 090902B and GRB 090926A (in this order) with $2\sigma$ error bars.
        	This correlation allows one to estimate off-axis angles of observed bursts.
        }
        \label{fig:correlations}
\end{figure}

The are some caveats and possible directions of improvement of the
model. First of all, as it's seen in Fig. \ref{fig:correlations} the
stretching factor of GRB 090926A is higher that the model
prediction. More data are needed to understand whether this requires a
refinement of the model or is a statistical fluctuation.

Second, the shapes of light curves produced by the model (see
fig. \ref{fig:sampleLightCurves}) differ considerably from observed
ones (e.g. fig. \ref{fig:grb090902B}). In particular, the slope of
light curves produced by the model is zero at $t = 0$, while the slope
of the observed light curves at $t = 0$ appears to have a maximum.

Third, the assumption of plasma energy dependence on off-axis angle
should be better justified.  This might probably be done using
hydrodynamic simulation of the jet.

We note, that even though it appears that bursts are similar in their
rest frames, this may be considered only as a first approximation.
The differences in masses, metallicities, angular momenta of
progenitor stars, and the differences in local conditions around them
might account for differences in light curves of GRBs (or, even, for
especially large stretching factor of GRB 090926A).  This means that
the minor variations of the model parameters might be necessary.

It seems, however, that all the issues above may be addressed by
refining the model without changing it's main assumptions.

Our study can be summarized with 3 main conclusions.

First, the time stretching of GRB light curves between different VHE
(in particular $\unit[100]{MeV} < E < \unit[1]{GeV}$ and $E >
\unit[1]{GeV}$) bands is discovered with the statistical significance
of $3.3\sigma$.  Depending on the burst the stretching factor may be
higher or lower than 1, that is the higher or the lower-energy light
curve is stretched.

Second, the time stretching may be explained with curvature effects,
that is the effects of jet geometry.  There is no need to introduce
any new spectral components.

Finally, one may assume that all GRBs are the same in their rest
frames.  The internal burst parameters (such as $\gamma$, $\eta_0$,
$r_0$, $n$, $\omega_0$, $\theta_0$, $k$ and $\alpha$) might stay the
same for all bursts, and this assumption is more or less consistent
with existing observations.

We note, that the model predicts a correlation between
the fraction of the high energy photons, the stretching factor and the
observer's off-axis angle (see fig. \ref{fig:correlations}). This may
be used as a method to estimate the observer's off-axis angle.

{\bf Acknowledgments.} We thank A.~Gruzinov, M.~Pshirkov and
S.~Troitsky for comments and inspiring discussions.  The work was
supported by Russian Science Foundation grant 14-12-01340. The
analysis is based on data and software provided by the Fermi Science
Support Center (FSSC). The numerical part of the work was done at the
cluster of the Theoretical Division of INR RAS. G.R. acknowledges the
fellowship of the Dynasty foundation.

\begin{appendices}

\section{Details of using Fermi Science tools}
\label{sec:fermiCode}
This appendix contains the details on using the Fermi-LAT
data. Namely, the exact values passed to the web form on the Fermi-LAT
data server and the code used to run the Fermi tools including the
full list of options.

Parameters that are passed to the web form of the Fermi-LAT data server to download the bursts data are the following:
\begin{itemize}
	\item{
		{\bf Object name or coordinates}.
		Coordinates are filled from the Table \ref{tab:bursts}.
	}
	\item{
		{\bf Coordinate system}.
		J2000.
	}
	\item{
		{\bf Search radius (degrees)}.
		$60$.
		Events are filtered by location separately by using the point spread functions (PSFs) of the LAT (as described in Section \ref{sec:photonSelection}).
	}
	\item{
		{\bf Observation dates}.
		To have a decent safety margin, we extend the duration of the burst by $50\%$ to both past and future relative to the Table \ref{tab:bursts} time ranges.
		So, we fill in the following values:
		\begin{align*}
			\texttt{time} + \texttt{startOffset} &- 0.5\left(\texttt{endOffset}-\texttt{startOffset}\right),\\
			\texttt{time} + \texttt{endOffset} &+ 0.5\left(\texttt{endOffset}-\texttt{startOffset}\right)
		\end{align*}
	}
	\item{
		{\bf Time system}.
		MET.
	}
	\item{
		{\bf Energy range (MeV)}.
		$100,\,300000$.
		This includes our both energy ranges.
	}
	\item{
		{\bf LAT data type}.
		Extended.
	}
	\item{
		{\bf Spacecraft data}.
		Checked.
		It is required for both events filtering, and calculation of the PSFs and exposure maps.
	}
\end{itemize}

For background, all the parameters are the same, except for the time ranges, which are the following:
\begin{align*}
	\texttt{time} + \texttt{startOffset} &- 0.5\left(\texttt{endOffset} - \texttt{startOffset}\right) - 86400,\\
	\texttt{time} + \texttt{startOffset} &- 0.5\left(\texttt{endOffset} - \texttt{startOffset}\right)
\end{align*}

The commands we use to run the Fermi Science Tools are the following:
\begin{lstlisting}
gtselect infile=[eventFile] outfile=filtered.fits
	ra=INDEF dec=INDEF rad=180 tmin=INDEF tmax=INDEF emin=100 emax=300000
	zmax=100 evclass=2 convtype=-1 evtable=EVENTS
\end{lstlisting}
Here \texttt{[eventFile]} is the file downloaded from the Fermi-LAT data server.

\begin{lstlisting}
gtmktime scfile=[spacecraft] sctable=SC_DATA
	filter="DATA_QUAL>0 && LAT_CONFIG==1" roicut=yes
	evfile=filtered.fits evtable=EVENTS outfile=timed.fits
	apply_fiter=yes
\end{lstlisting}
where \texttt{[spacecraft]} is the FITS file containing the spacecraft data (also downloaded from the Fermi-LAT data server).

\begin{lstlisting}
gtltcube evfile=timed.fits evtable=EVENTS scfile=[spacecraft] sctable=SC_DATA
	outfile=ltcube.fits
	dcostheta=0.025 binsz=1 phibins=0 tmin=0 tmax=0 zmax=100 zmin=0
\end{lstlisting}

\begin{lstlisting}
gtpsf expcube=ltcube.fits outfile=psf_[IrfName].fits outtable=PSF irfs=[IrfName]
	ra=[RA] dec=[DEC] emin=100 emax=300000 nenergies=41 thetamax=30 ntheta=300
\end{lstlisting}
\texttt{[RA]} and \texttt{[DEC]} here are the coordinates of the
burst, and \texttt{[IrfName]} is the Instrument Response Function,
which depends on photon's event class (``SOURCE'' in our case) and
conversion type (back or front). The list of IRF names is obtained
with the \texttt{gtirfs} tool.

\begin{lstlisting}
gtexpcube2 infile=ltcube.fits cmap=none outfile=expcube_[IrfName].fits irfs=[IrfName]
	nxpix=360 nypix=180 binsz=1 coordsys=CEL xref=0 yref=0 axisrot=0
	proj=CAR ebinalg=log emin=100 emax=300000 enumbins=40
	ebinfile=NONE bincalc=EDGE ignorephi=no thmax=180 thmin=0 table=Exposure
\end{lstlisting}
Note, that R.A. coordinates in the \texttt{expcube} FITS files are
indexed in reverse order and start from 180. For example,
$\texttt{ra[0]}=180$, $\texttt{ra[1]}=179$, and so on.

\section{Cosmological distances}
\label{sec:cosmology}
This appendix contains a few formulas related to the cosmological
distances. In particular, it provides formula for the area of a photon
sphere $A_\text{ph}\left(z\right)$, a sphere over which photons
emitted in a particular burst at redshift $z$ are spread; and the
volume of the infinitesimal shell surrounding a burst sphere $\dif
V\left(z\right)$, a sphere over which bursts at redshift $z$ are
distributed.

First of all, the metric of the expanding Universe:
\begin{equation}
\dif s^2 = -\dif t^2 + a^2\left(t\right) \dif r^2 + a^2\left(t\right) r^2 \dif \Omega_2
\end{equation}
We define $a\left(t_\text{obs}\right) = 1$, where $t_\text{obs}$ is
the observation time.

We need to understand how the scale factor changes with time. For that we assume that the energy content of the Universe consists of matter $\Omega_m$ and vacuum energy $\Omega_\Lambda$ only, so that the Friedmann equation takes the following form:
\begin{align*}
\left( \frac{\dot{a}\left(t\right)}{a\left(t\right)} \right)^2 &= \Omega_m H_\text{obs}^2 \frac{1}{a^3\left(t\right)} + H_\text{obs}^2 \Omega_\Lambda \\
\dot{a}\left(t\right) &= a\left(t\right) H_\text{obs} \sqrt{\Omega_m\frac{1}{a^3\left(t\right)} + \Omega_\Lambda} \\
\dif t &= \frac{\dif a}{a\left(t\right) H_\text{obs} \sqrt{\Omega_m\frac{1}{a^3\left(t\right)} + \Omega_\Lambda}}
\end{align*}
Here $H_\text{obs} = H\left(t_\text{obs}\right) = \frac{\dot{a}\left(
  t_\text{obs} \right)}{a\left( t_\text{obs} \right)} = \dot{a}\left(
t_\text{obs} \right)$ is the Hubble parameter at the observation time.

We also need to know the areas of two spheres: the photon sphere, a
sphere over which photons emitted in a particular burst are spread;
and the bursts sphere, a sphere over which bursts at a particular
redshift are distributed. These two spheres have the same radii -- a
distance from the observer to the burst, but different centers: the
photon sphere is centered on the burst, while the bursts sphere is
centered on the observer. Also, they have different areas, because the
scale factor differs at the time of emission and observation.

Let us begin with the photon sphere. This sphere has the origin at $r = 0$ and $t = 0$, at the central engine of a particular burst. Taking $\dif s = 0$ and $\dif \Omega_2 = 0$ in the equation for metric, we get:
\begin{equation*}
\dif r = \frac{\dif t}{a\left(t\right)}
\end{equation*}
And now we integrate over time to find the observer's position:
\begin{align*}
r\left(t_\text{obs}\right) &= \int_0^{t_\text{obs}}\frac{\dif t}{a\left(t\right)} \\
&= \int_{a\left(0\right)}^1\frac{\dif a}{a^2\left(t\right) H_\text{obs} \sqrt{\Omega_m\frac{1}{a^3\left(t\right)} + \Omega_\Lambda}} \\
&= \frac{\, _2F_1\left(\frac{1}{3},\frac{1}{2};\frac{4}{3};-\frac{\Omega_m}{a^3\left(0\right) \Omega_\Lambda }\right)-a\left(0\right) \, _2F_1\left(\frac{1}{3},\frac{1}{2};\frac{4}{3};-\frac{\Omega_m}{\Omega_\Lambda }\right)}{a\left(0\right) H_\text{obs} \sqrt{\Omega_\Lambda }}
\end{align*}
Here $_2F_1\left(a, b; c; z\right) =
\frac{\Gamma\left(c\right)}{\Gamma\left(b\right)\Gamma\left(c-b\right)}\int_0^1
\frac{t^{b-1}\left(1-t\right)^{c-b-1}}{\left(1-t z\right)^a} \dif t$
is a hypergeometric function. We then substitute $a\left(0\right) =
\frac{1}{1+z}$ and finally arrive at
\begin{equation}
r\left(z\right) = \frac{\left(1+z\right)\, _2F_1\left(\frac{1}{3},\frac{1}{2};\frac{4}{3};-\frac{\Omega_m}{\Omega_\Lambda}\left(1+z\right)^3\right) - \, _2F_1\left(\frac{1}{3},\frac{1}{2};\frac{4}{3};-\frac{\Omega_m}{\Omega_\Lambda }\right)}{H_\text{obs} \sqrt{\Omega_\Lambda }}
\end{equation}
The area of the photon sphere is then:
\begin{equation}
A_\text{ph}\left(z\right) = 4 \pi a^2\left(t_\text{obs}\right) r^2\left(z\right) = 4 \pi r^2\left(z\right)
\end{equation}

The second sphere is the bursts sphere. It has its origin at the
observer's position, and the radius the same as the photon
sphere. Therefore, its area is:
\begin{equation}
A_\text{b}\left(z\right) = 4 \pi a^2\left(0\right) r^2\left(z\right) = \frac{4 \pi r^2\left(z\right)}{\left(1+z\right)^2}
\end{equation}

It's helpful to calculate one more quantity related to the bursts
sphere -- the volume of the infinitesimal shell surrounding it. For
that we again use the metric:
\begin{align}
\dif V\left(z\right) &= - A_\text{b}\left(z\right) a\left(0\right) \dif r \nonumber\\
&= - A_\text{b}\left(z\right) a\left(0\right) \frac{\dif a}{a^2\left(0\right) H_\text{obs} \sqrt{\Omega_m\frac{1}{a^3\left(t\right)} + \Omega_\Lambda}} \nonumber\\
&= A_\text{b}\left(z\right) \frac{\frac{1}{\left(1+z\right)^2}\dif z \left(1+z\right)}{H_\text{obs} \sqrt{\Omega_m\frac{1}{a^3\left(t\right)} + \Omega_\Lambda}} \nonumber\\
&= \frac{4\pi r^2\left(z\right)}{\left(1+z\right)^3} \frac{\dif z}{H_\text{obs} \sqrt{\Omega_m\left(1+z\right)^3 + \Omega_\Lambda}}
\end{align}

\end{appendices}

\bibliographystyle{hieeetr}
\bibliography{gammaRays.bib}

\end{document}